\documentclass[10pt]{iopart}

\usepackage{graphicx}
\usepackage{subfigure}
\usepackage{amsfonts}
\expandafter\let\csname equation*\endcsname\relax
\expandafter\let\csname endequation*\endcsname\relax
\usepackage{amsmath}
\usepackage{enumerate}
\usepackage{comment}
\usepackage{color}
\usepackage{citesort}
\newcommand{\bA}{\mathbf{A}}

\newcommand{\calN}{\mathcal{N}}
\newcommand{\calM}{\mathcal{M}}

\newcommand{\bk}{\mathbf{k}}

\newcommand{\Order}[1]{{\mathcal{O} \left( #1 \right)}}
\newcommand{\s}[1]{\left[ #1 \right]}
\newcommand{\expected}[1]{\left\langle #1 \right\rangle}
\newcommand{\p}[1]{\left( #1 \right)}

\begin{document}

\title[Loopy random graphs with fixed degrees and arbitrary degree distributions]{Transitions in loopy random graphs with fixed degrees and arbitrary degree distributions}

\author{Fabi\'an Aguirre L\'opez$^{1,2}$ and Anthony CC Coolen$^{3,4}$}
\address{$^1$Dept of Mathematics, King's College London, Strand, London WC2R2LS, UK\\ 
$^2$Inst for Mathematical and Molecular Biomedicine, King's College London, Hodgkin Building, London SE11UL, UK\\
$^3$Dept of Biophysics, Radboud University, 6525AJ Nijmegen, The Netherlands\\
$^4$London Inst for Mathematical Sciences, 35A South St, London W1K2XF, UK
}

\ead{fabian.aguirre\_lopez@kcl.ac.uk, a.coolen@science.ru.nl}

\begin{abstract}
We analyze maximum entropy random graph ensembles with constrained degrees, drawn from arbitrary degree distributions, and a tuneable number of 3-loops (triangles). We find that such ensembles generally  exhibit two transitions, a clustering and a shattering transition, separating three distinct regimes. At the clustering transition, the graphs change from typically having only isolated loops to forming loop clusters. At the shattering transition the graphs break up into extensively many small cliques to achieve the desired loop density.  The locations of both transitions depend nontrivially on the system size. We derive a general formula for the loop density in the regime of isolated loops, for graphs with degree distributions that have finite and second moments. For bounded degree distributions we present further analytical results on loop densities and phase transition locations, which, while non-rigorous, are all validated via MCMC sampling simulations. We show that the shattering transition is  of an entropic nature, occurring for all  loop density values, provided the system is large enough.
\end{abstract}

\pacs{64.60.aq, 02.10.Ox, 64.60.De}% PACS, the Physics and Astronomy 
% Classification Scheme.
%64.60.aq Networks
%02.10.Ox Combinatorics; graph theory
%64.60.De Statistical mechanics of model systems (Ising model, Potts model, field-theory models, Monte Carlo techniques, etc.)
%\keywords{random graph ensembles, clustering, cycles, transitivity}
\maketitle

% \tableofcontents

\section{Introduction}

Graph theory was introduced by Euler to solve the problem of the seven bridges of K\"onigsberg \cite{euler1741solutio}. He noted that upon stripping all unnecessary details to solve this problem, one was left only with a set of $4$ nodes and  $7$ links between them. Since then, networks and graphs have proven to be fundamental in the modelling of many real world phenomena.  While 
 with the advent of powerful computers accessible to almost all researchers it is now typical for network scientists to work on a daily basis with networks of nodes ranging from thousands to millions, still the modelling strategy is the same: remove unnecessary details and reduce the problem to nodes and links.
 
For scientists, and especially those with a statistical training -- used to thinking in terms of null models in hypothesis testing \cite{casella2002statistical} -- it is natural to ask a very simple question regarding observed networks: which are typical and which are atypical topological features? To answer this question  one commonly works with  random graph ensembles, designed  to mimic real-world networks; see e.g. \cite{solomonoff1951connectivity,erdHos1960evolution,annibale2017generating,newman2003structure}. In addition to studying  properties of graphs, one usually also seeks to understand processes for which these graphs define the interaction infrastructure, and the relation between graph topology and process efficacy.  Here one would benefit from exact analytical solutions for processes defined on nontrivial graph ensembles. Unfortunately, this is hard. While  there has been an explosion of exact solutions for processes on random graphs, the vast majority of these are locally tree-like graphs. This property allows one to write recursive equations, that become exact for large graphs and show very good agreement with simulations on finite ones. Ironically, this property that makes the models solvable is the same property that makes them unrealistic. 

In addition to the previous complication, there is also the fact that there is no easily controllable  random graph ensemble that generates graphs with given numbers of links and  triangles. The natural extension of the Erd\"{o}s-R\'{e}nyi ensemble \cite{ER}  was presented by Strauss  \cite{strauss1986general}, who observed that the ensemble condensed very quickly into dense graphs, losing any resemblance to real networks. There is a long history of attempts at understanding this transition \cite{jonasson1999random,burda2004network,chatterjee2013estimating,yin2016detailed}, and many alternative loopy random graph ensembles and algorithms have since then been presented \cite{holme2002growing,guo2009generator,newman2009random,miller2009percolation,foster2010communities,bianconi2014triadic,tamm2014islands,lopez2018exactly,lopez2018exactly}, yet none generate easily controllable graphs. It appears very hard to access a regime where there is high number of triangles while keeping a `nice'  topology. More recent  models conserve the degree sequence to avoid the condensation observed by Strauss, but still show a transition into a clustered regime \cite{tamm2014islands,avetisov2016eigenvalue,lopez2018exactly,avetisov2018phase,pospelov2019spectral,avetisov2019localization,lopez2020imaginary}. The logical way of stopping the appearance of a clustered regime is to restrict the number of triangles in each node via a hard constraint, as proposed in \cite{newman2009random,miller2009percolation}. While there have been numerical and theoretical advances with this model \cite{hackett2011cascades,volz2011effects,herrero2015ising,peron2018spectra,herrero2019self,cantwell2019message},  it remains difficult to keep the target degree distribution and the target total number of triangles under control  \cite{heath2011generating}. 

In this paper we study a random graph ensemble with a tuneable number of loops, achieved with a soft global constraint on the number of triangles in combination with a hard constraint on the degrees, each drawn from a fixed degree distribution. This guarantees that our graphs will both be loopy and sparse, which are desired properties to mimic real networks. 
This model was previously studied in \cite{foster2010communities}. However, in that previous study  the chosen MCMC move acceptance probabilities did not ensure convergence to the target distribution. This was pointed out in \cite{sampling,annibale2017generating}, where it was shown that in edge swap graph dynamics nontrivial acceptance probabilities are needed (see also \ref{sec:numerical-sampling}).  
We show in this paper that with the correct MCMC sampling the model again displays a transition into a clustered phase, and that the overall phenomenology presented in \cite{foster2010communities} coincides with our results.
We then proceed to develop an extended theoretical and quantitative understanding of the behaviour of the model, including an analytic characterization of the low triangle density phase, expressions for the locations of the two (clustering and shattering) transitions, and scalar measures to probe the interactions between loops and their relevance for the phases of the ensemble. Our results and predictions are supported via nontrivial graph sampling simulations involving  different degree distributions, using the exact move acceptance probabilities of \cite{sampling,annibale2017generating}.

\section{The model}

We study a random graph ensemble defined on the set of $N$-node graphs with a given degree distribution. A graph is an ordered pair $(V,E)$ of nodes and edges, respectively. We model graphs through their adjacency matrices $\bA$, defined by the entries $A_{ij} = 1$ if $(i,j)\in E$, and $0$ otherwise. We will only be concerned with simple undirected graphs, which in terms of the adjacency matrix implies that $A_{ij}=A_{ji}$ and $A_{ii}=0$ for all $(i,j)$. The degree of a node is the number of edges connected to it, $k_i(\bA) = \sum_{j}A_{ij}$. Throughout this paper we will work with graphs that have exactly the same degree sequence $\{k_i\}_{i=1,\dots,N}$. Each element of this sequence is drawn randomly and independently from a given distribution $p(k)$. In the large $N$ limit we know that the empirical distribution of degrees will converge to the target distribution, $p(k) = \lim_{N\to\infty }N^{-1}\sum_{i=1}^N \delta_{k,k_i}$.

The number of triangles in a graph is easy to calculate if  we identify them with the loops of length three, up to overcounting. There is a loop of length three around node $i$ if there exist $j$ and $k$ such that $(i,j)\in E$, $(j,k)\in E$, and $(k,i)\in E$. Since our graph is simple, the indices $i,j,k$ are all different. In the language of the adjacency matrix, the indicator function for a given 3-loop takes the simple form
\begin{equation}
    \mathbb{I} \s{(i\to j\to k \to i) \in \bA } = A_{ij} A_{jk} A_{ki}.
\end{equation}
The total number of 3-loops is then simply the trace of the third power of the adjacency matrix (modulo overcounting by a factor 6),
\begin{align}
    \calM(\bA) = \sum_{ijk}\mathbb{I} \s{(i\to j\to k \to i) \in \bA } = \Tr(\bA^3).
\end{align}
We now define an ensemble of random graphs such that the average number of triangles can be controlled, using a parametrized distribution over graphs denoted by $p(\bA)$. Our choice is a maximum entropy (ME) ensemble. That is, we take $p(\bA)$ to be such that the average number of triangles is fixed,
\begin{align}
    \calM^* = \sum_{\bA} p(\bA)\Tr(\bA^3),
\end{align}
and that the degree sequence $\bk=\{k_i\}_{i=1,\dots,N}$ is achieved exactly. Among those distributions $p(\bA)$ that share these two properties, we choose the one that maximizes the Shannon entropy $S[p] = -\sum_{\bA}p(\bA)\log p(\bA)$. This will guarantee that the distribution is statistically unbiased \cite{infotheory}.
The ME distribution is of an exponential form, with one tuneable parameter $\alpha$,
\begin{align}
\label{eq:ensemble}
    p(\bA) = \frac{1}{Z(\alpha)}\rme^{\alpha \Tr (\bA^3)}\prod_{i=1}^N\delta_{k_i,\sum_j A_{ij}}.
\end{align}
The product over Kronecker deltas enforces the degree sequence of the graph. For $\alpha =0 $ the ensemble reduces to the configuration model \cite{bollobas1980probabilistic} (CM), a uniform distribution over all graphs with degree sequence $\bk$.

Our main observable of the ensemble will be the number of 3-loops \emph{per node}. We will refer to it as the \emph{loop density},
\begin{align}
    m(\alpha) =N^{-1} \expected{\calM(\bA)} = N^{-1}\expected{\Tr(\bA^3)}.
\end{align}
Where $\expected{f(\bA)} = \sum_{\bA}p(\bA)f(\bA)$. This quantity reflects the typical number of loops in the neighborhood of  a node. Each node can have a different maximum number of triangles, depending on its degree.
Once a random graph ensemble like (\ref{eq:ensemble}) is defined it is desirable to have both an algorithm to generate graph samples numerically and an analytic theory of its statistical properties.
In order to generate samples form an ensemble such as (\ref{eq:ensemble}) we use a Markov Chain Monte Carlo (MCMC) approach. The algorithm starts with a seed graph satisfying the degree sequence, and evolves it by performing degree preserving edge swaps as shown in Figure \ref{fig:edgeswap}. Edge swaps are either accepted or rejected, with a nontrivial acceptance probability that not only takes into account the specific ensemble (\ref{eq:ensemble}) but also the availability of possible edge swaps as the graph evolves. The theory of this MCMC algorithm was developed in \cite{sampling} and presented more extensively in \cite{annibale2017generating}. It is also summarized briefly in \ref{sec:numerical-sampling}.

To find an analytic expression for the loop density we need to calculate the  generating function $\phi(\alpha)$:
\begin{align}
\label{eq:freeEnergy}
    \phi(\alpha) &= \frac{1}{N}\log Z(\alpha) = \frac{1}{N} \log \sum_{\bA} \rme^{\alpha \Tr (\bA^3)} \prod_{i=1}^N \delta_{k_i ,\sum_j A_{ij}} \\
    \label{eq:malpha}
    m(\alpha) &= \frac{\partial\phi(\alpha)}{\partial\alpha} = \expected{\frac{1}{N}\Tr(\bA^3)}
\end{align}
Ideally, knowledge of the functions $\phi(\alpha)$ and $m(\alpha)$ would allow us to generate random graphs with any desired loop density. Although it is not possible to calculate $\phi(\alpha)$ analytically, in section \ref{sec:analytic} we will show that a small $\alpha$ approximation will give very good results for a wide range of values. Additionally we will give a description of the general behaviour of this ensemble for the whole range of $\alpha$ values.
Another important observable of the ensemble reports on the amount of \emph{interaction} between the triangles in the graph, i.e. the number of edges and nodes that different triangles share. This varies in a nontrivial way with different values of $\alpha$ and different system sizes $N$. 
To measure the degree of interaction between loops, we define 
\begin{align}
\label{eq:rdef}
    r(\bA) = \frac{\# \textrm{nodes in triangles}}{\# \textrm{ of triangles}} =  \frac{\sum_{i=1}^N \Theta[(\bA^3)_{ii}]}{\frac{1}{6}\Tr(\bA^3)}\in[0,3],
\end{align}
where $\Theta(x) = 1$ if $x>0$ and zero otherwise. This ratio of triangle vertices to triangles is independent of the total number of triangles in the graph. If $r(\bA)=3$, the triangles are all non-interacting in the sense that they do not share any nodes. If $r(\bA)<3$ triangles are sharing nodes. Some simple examples are shown on the top row of Figure \ref{fig:rExamples}. In the particular case where graphs form cliques of $q+1$ nodes, we would have $r(\bA) = 6/(q^2-q)$; this is a natural lower bound for graphs of maximum degree $q$.

\subsection{Main results}

We will now outline the main results of our analysis of the ensemble (\ref{eq:ensemble}).
We found the same initial behaviour for all degree distributions as the triangle-inducing control parameter $\alpha$ is increased form $\alpha = 0$. This behaviour depends only  on  the first two moments of the degree distribution, $c = \overline{k}$ and $\overline{k^2}$ (where $\overline{f(k)} = N^{-1}\sum_{i=1}^N f(k_i)$), and on the maximum degree $q = \max_{i=1,\dots,N}\{k_i\}$ (for bounded degree distribution). We will only consider the case where $N$ is sufficiently large,  $Np(q)\gg q\!+\!1$, so that all degrees are typically represented in the graph with an extensive number of nodes.

\begin{figure}[t]
    \centering
    \begin{picture}(340,285)
        \put(10,165){\includegraphics[width = 0.44\textwidth]{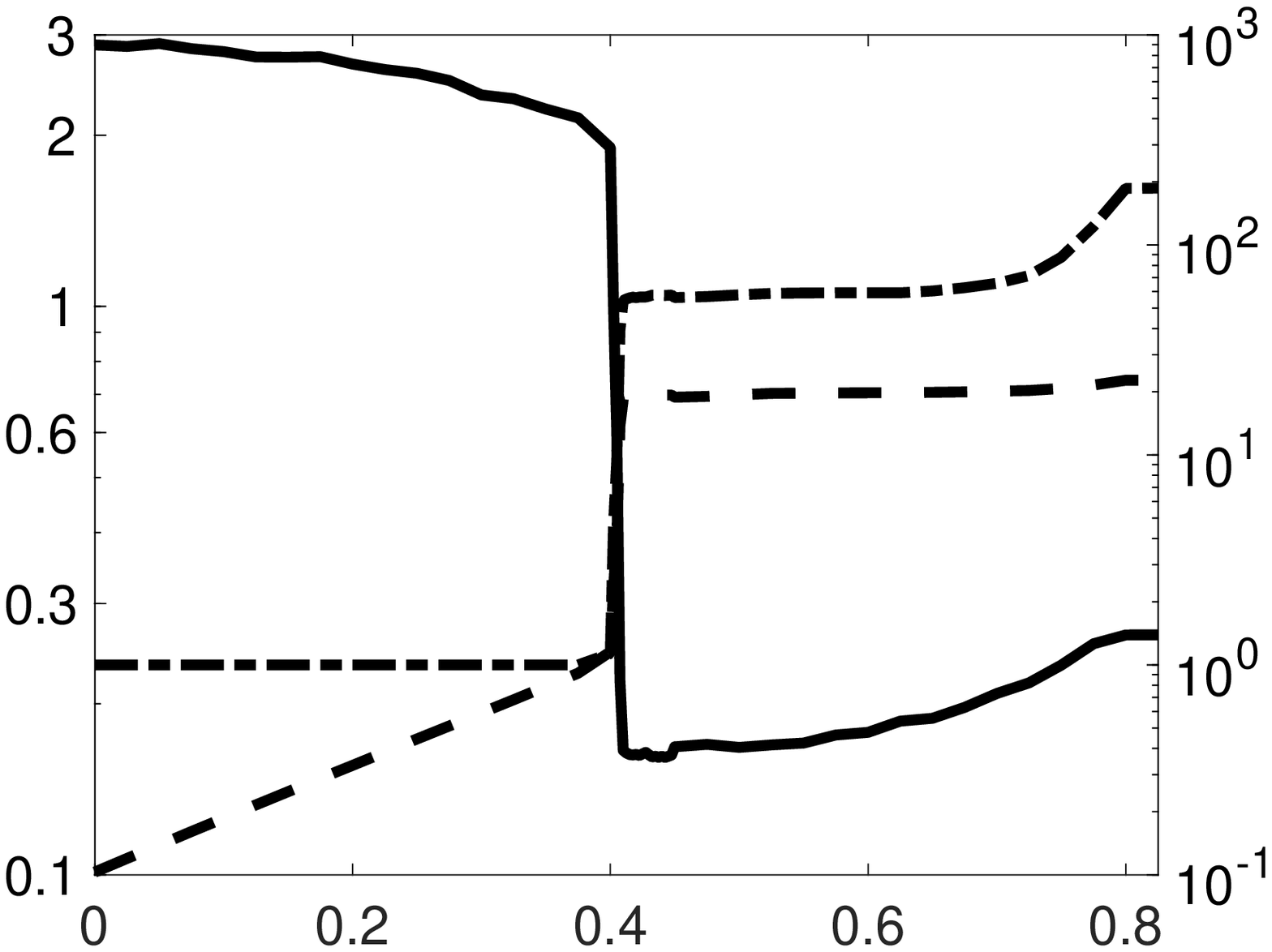}}
        \put(185,165){\includegraphics[width = 0.44\textwidth]{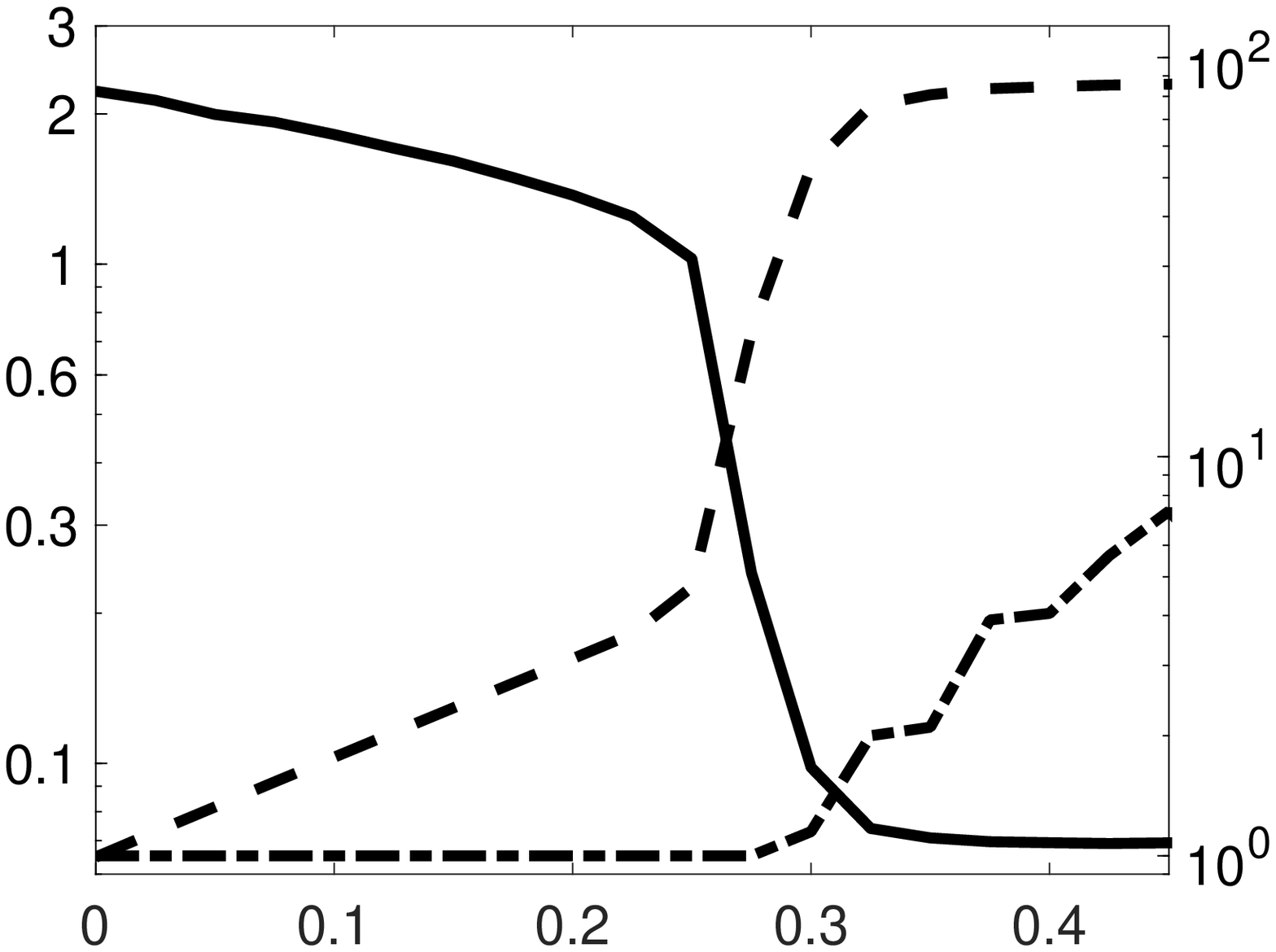}}
        \put(10,20){\includegraphics[width = 0.44\textwidth]{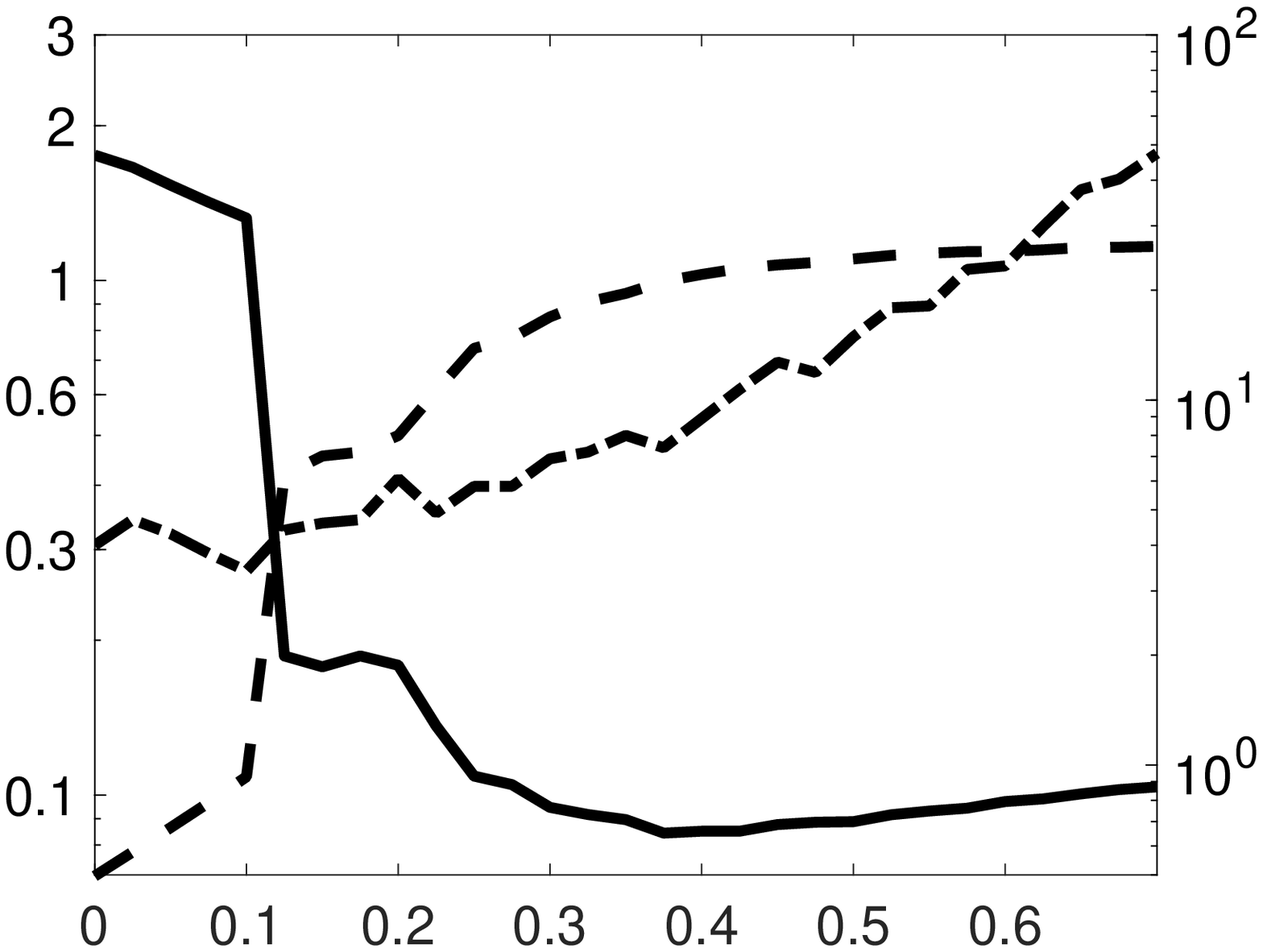}}
        \put(185,20){\includegraphics[width = 0.44\textwidth]{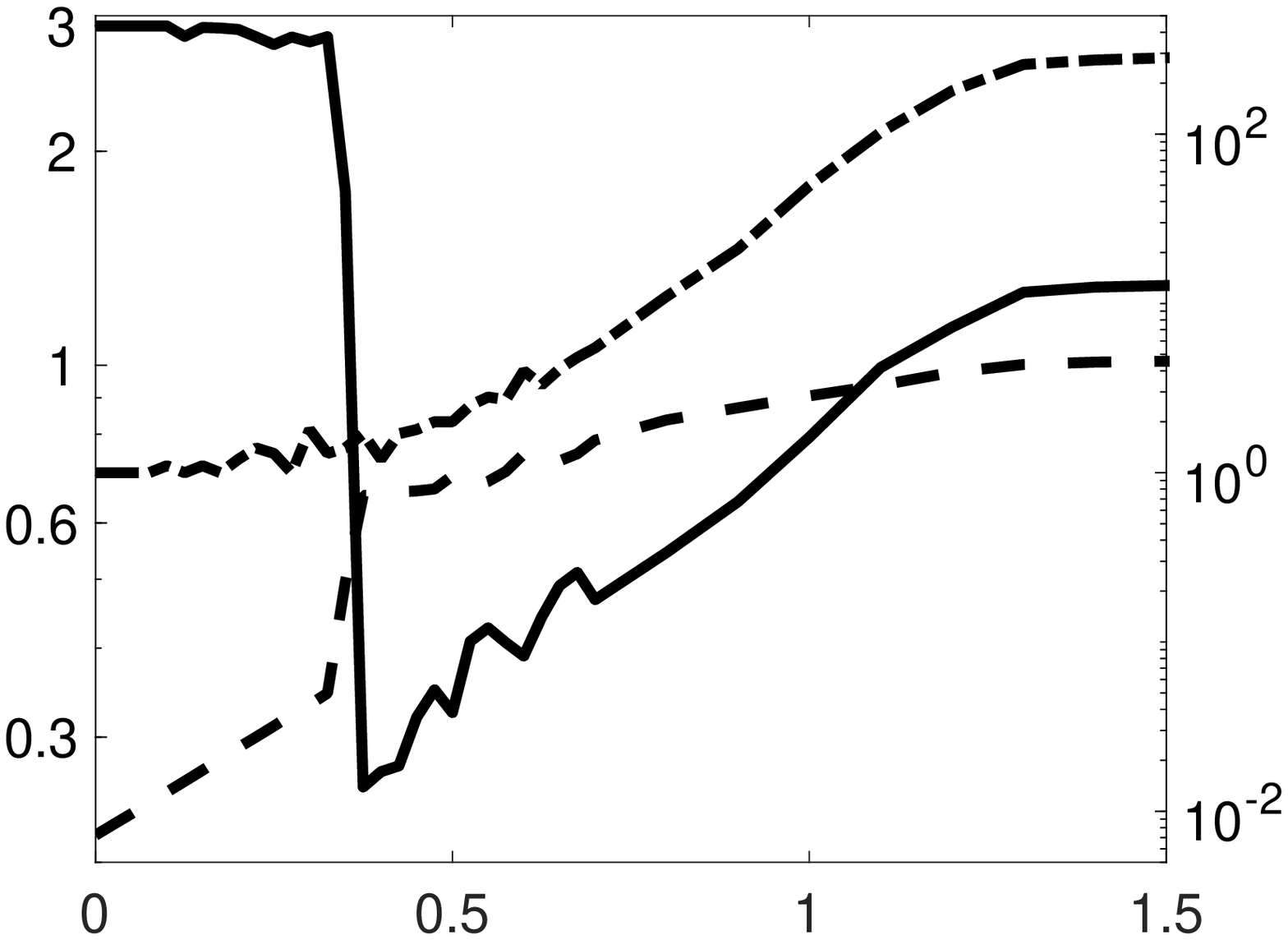}}
        \put(95,15){$\alpha$}
        \put(95,160){$\alpha$}
        \put(260,15){$\alpha$}
        \put(260,160){$\alpha$}
        \put(40,265){$\leftarrow r$}
        \put(130,260){$n\rightarrow$}
        \put(130,226){$m\rightarrow$}
        %\put(92,2){$(c)$}
        %\put(257,2){$(d)$}
        %\put(92,147){$(a)$}
        %\put(257,147){$(b)$}
    \end{picture}
    \vspace*{-5mm}
    \caption {
    Values of the three main topological observables as measured in numerical sampling simulations of the random graph ensemble (\ref{eq:ensemble}), all with $N=1000$. 
  The  loop interaction observable $r(\alpha)$, defined in  (\ref{eq:rdef}), is shown as solid lines with values on the left vertical axis. The triangle density  $m(\alpha)$ is  shown as  dashed lines with values on the right vertical axis. The number of connected graph components  $n(\alpha)$ is shown as dashed-dotted line with values on the  right vertical axis. Different panels refer to different degree distribution. Top left: $p(k) = bim(k|3,7)$; top right: $p(k) = Poiss(k|10)$; bottom left: $p(k) = exp(k|4)$; bottom right: $p(k) = PL(k)$. See Table   \ref{tab:degreeDistributions} for the corresponding definitions. Error bars are omitted to avoid clutter; see Figures \ref{fig:rExamples} and \ref{fig:transition} for examples of typical error bar values.}
    \label{fig:m-vs-alpha}
\end{figure}

\begin{figure}[t]
    \centering
    \begin{picture}(325,282)
        \put(0,155){\includegraphics[width = 0.39\textwidth]{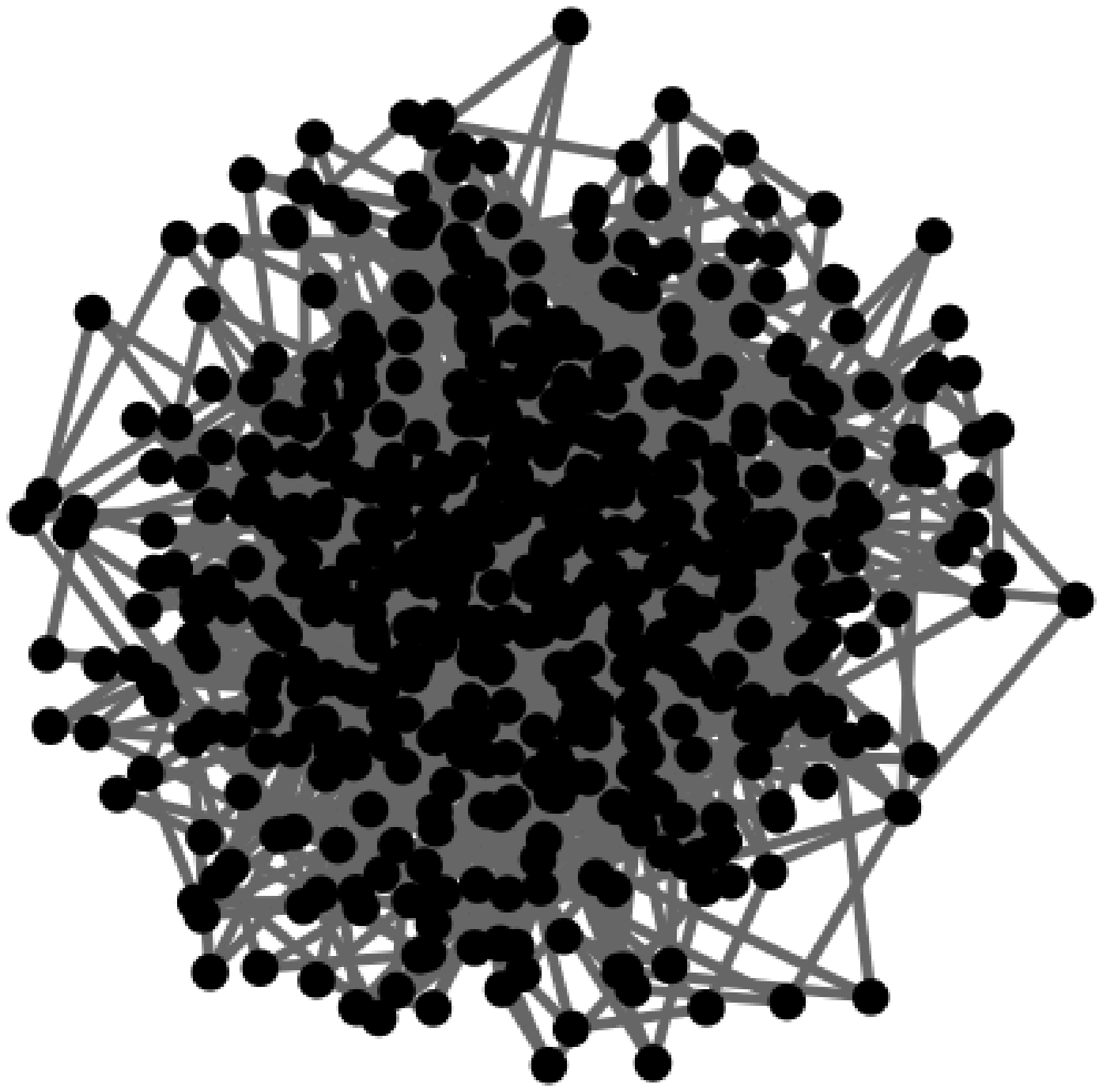}}
        \put(180,155){\includegraphics[width = 0.39\textwidth]{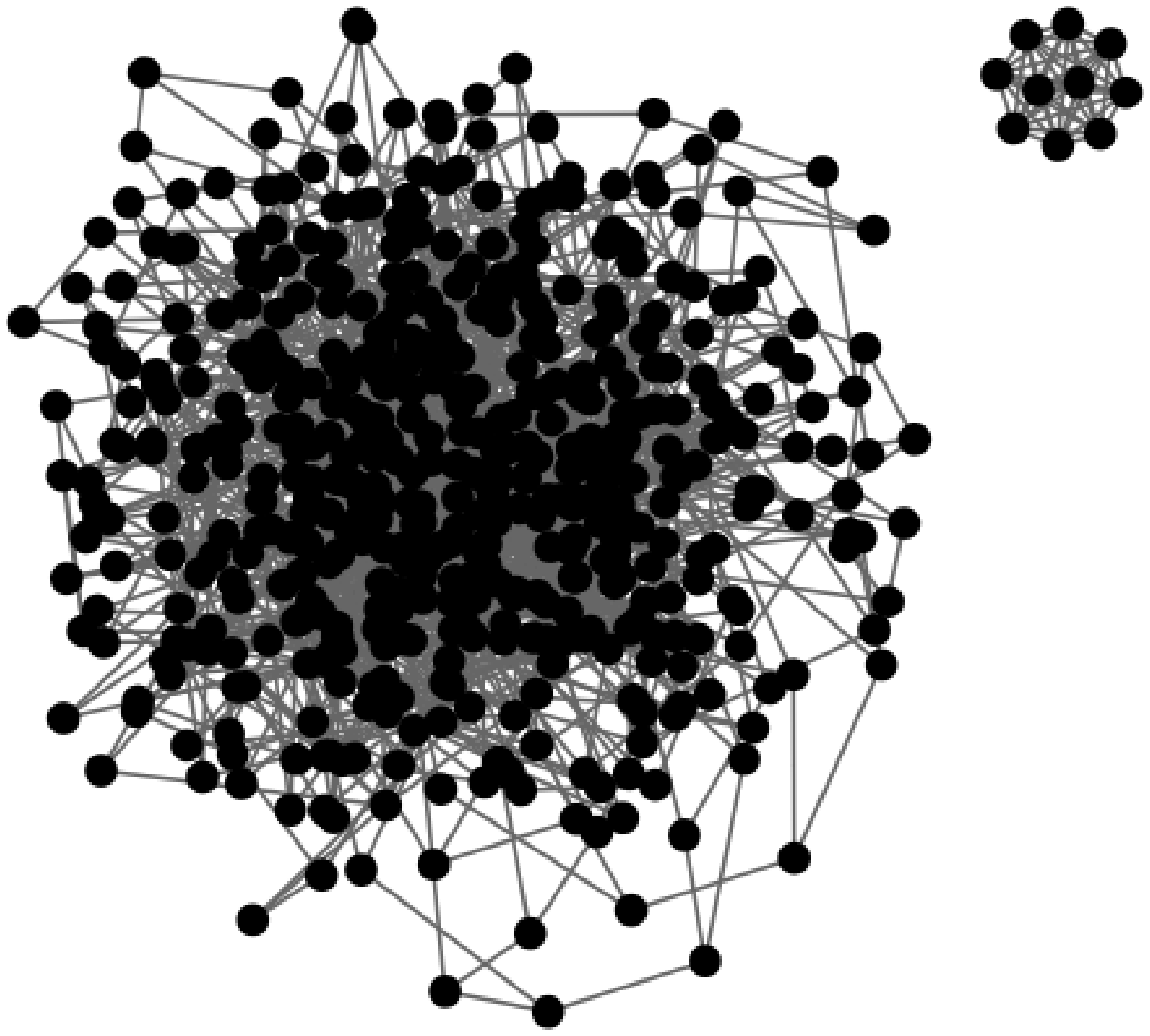}}
      \put(0,270) {$\alpha=0.125$:}  \put(0,258.5) {$m=1.1$}
         \put(180,270) {$\alpha=0.275$:}  \put(180,258.5) {$m=4.6$}
       
        \put(0,0){\includegraphics[width = 0.39\textwidth]{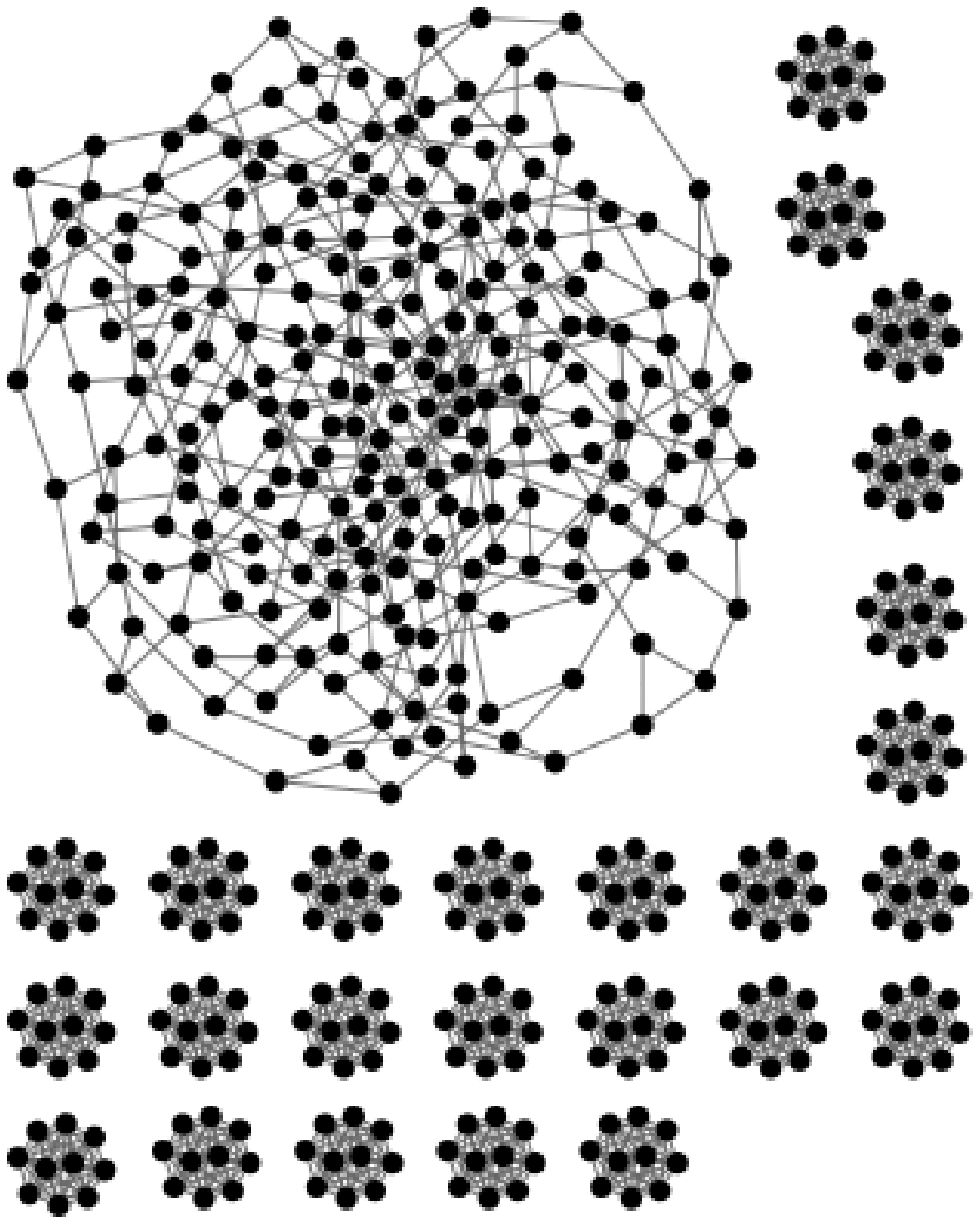}}
        \put(180,10){\includegraphics[width = 0.39\textwidth]{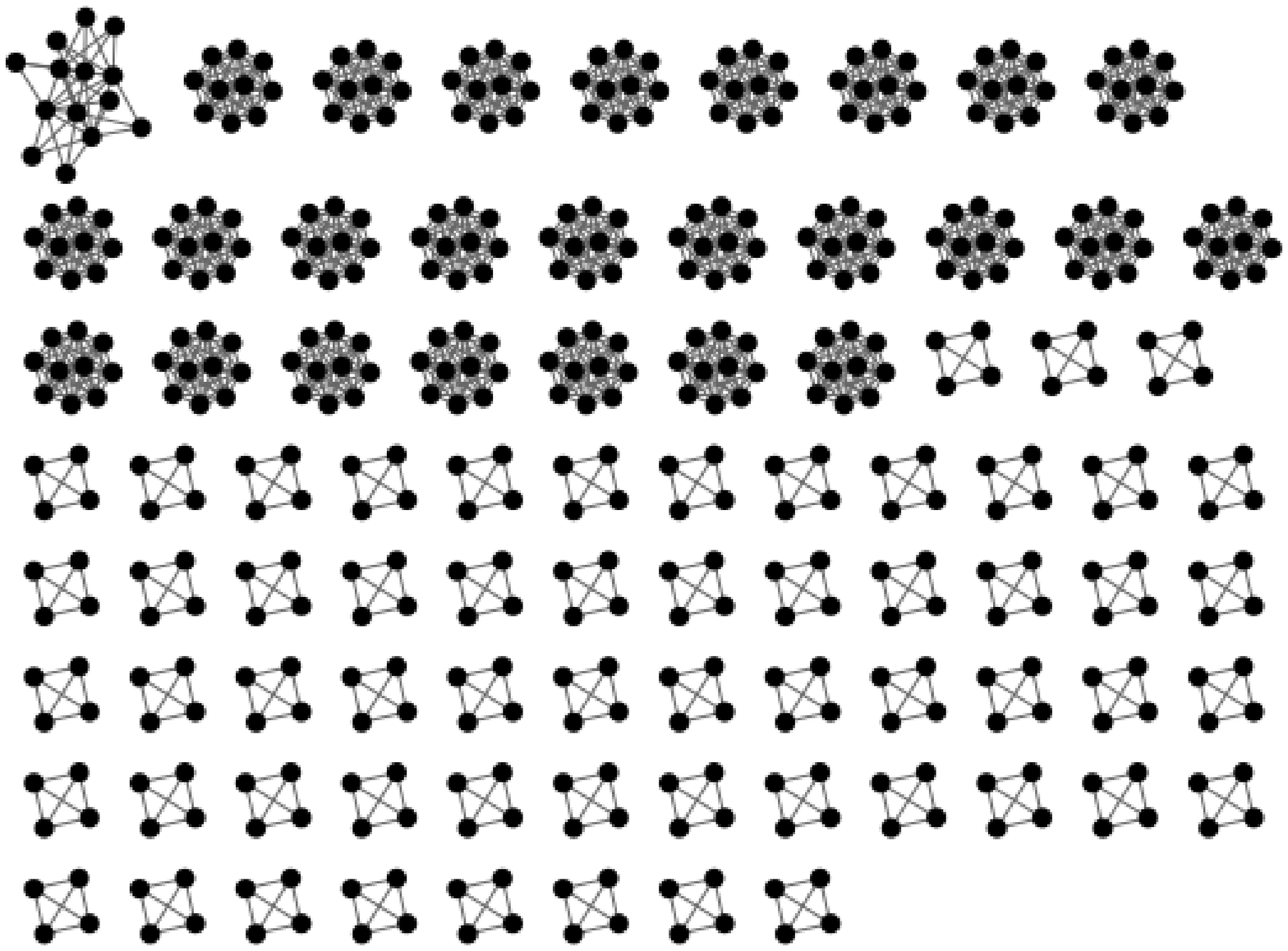}}  
            \put(0,140) {$\alpha=0.35$:}  \put(0,128.5) {$m=36.2$}
           \put(180,140) {$\alpha=0.75$:}  \put(180,128.5) {$m=39.4$}
                  
           \end{picture}
    \caption{
   Results of  numerical sampling simulations of the random graph ensemble (\ref{eq:ensemble}), with $N=500$ and $p(k) = \frac{1}{2}\delta_{k,3} + \frac{1}{2}\delta_{k,9}$. The four different images correspond to four different values of $\alpha$, with different loop densities $m(\alpha)$, as indicated. 
  }
    \label{fig:graph-examples-2}
\end{figure}

\begin{figure}[t]
    \centering
    \begin{picture}(325,282)
        \put(0,150){\includegraphics[width = 0.39\textwidth]{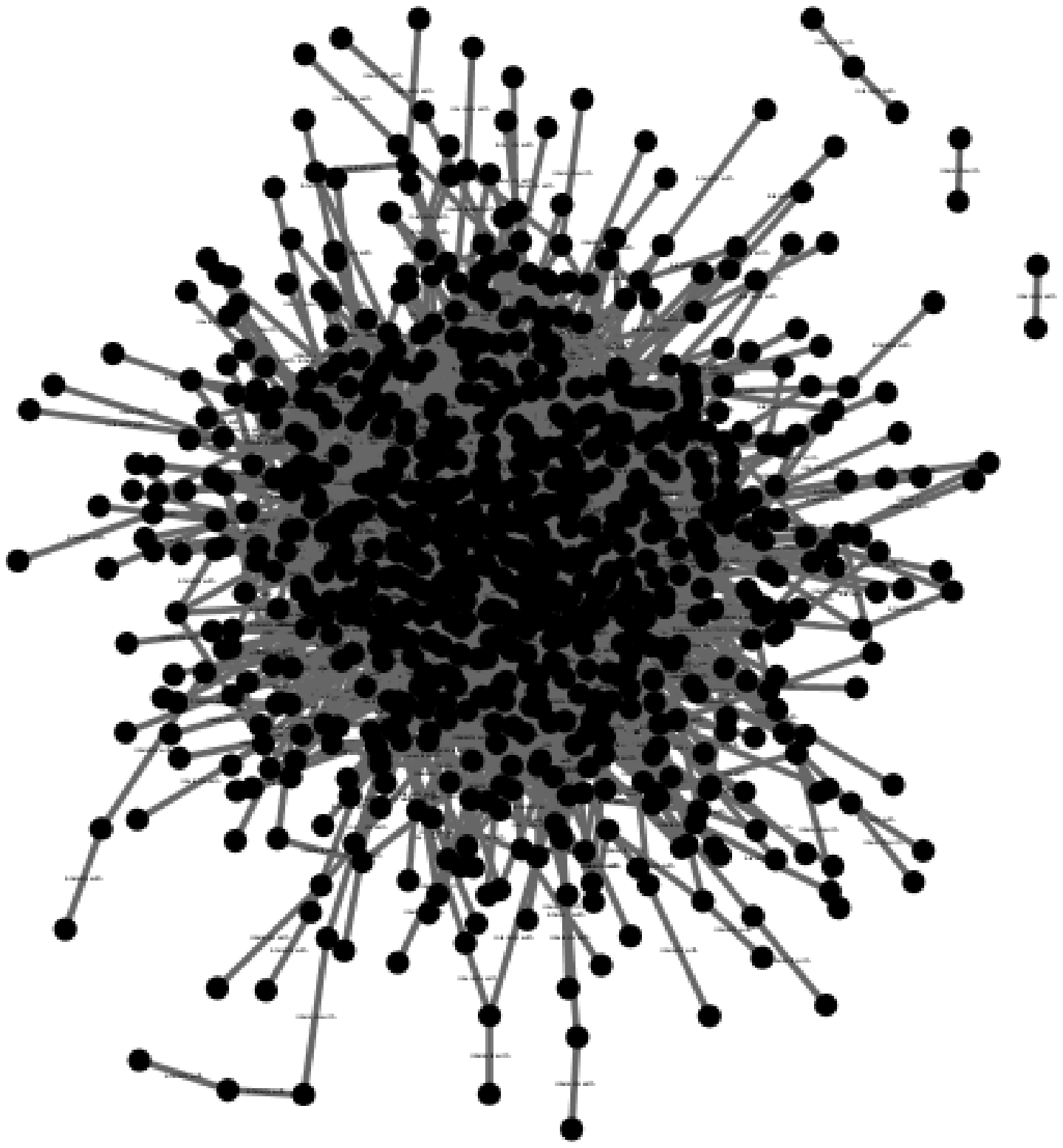}}
        \put(180,145){\includegraphics[width = 0.39\textwidth]{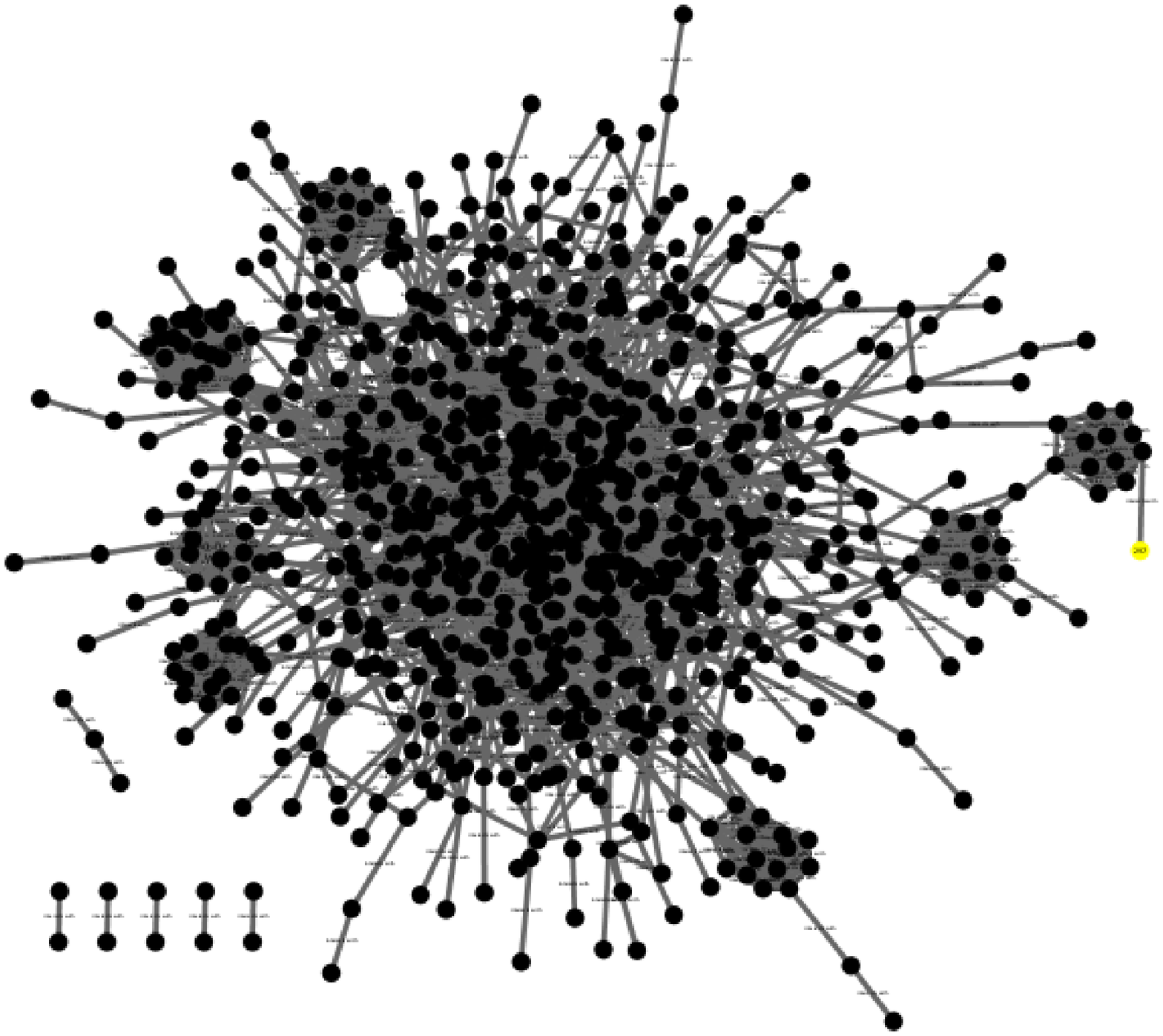}}
          \put(0,270) {$\alpha=0.075$:}  \put(0,258.5) {$m=0.8$}
         \put(180,270) {$\alpha=0.3$:}  \put(180,258.5) {$m=16.9$}

        \put(0,-3){\includegraphics[width = 0.39\textwidth]{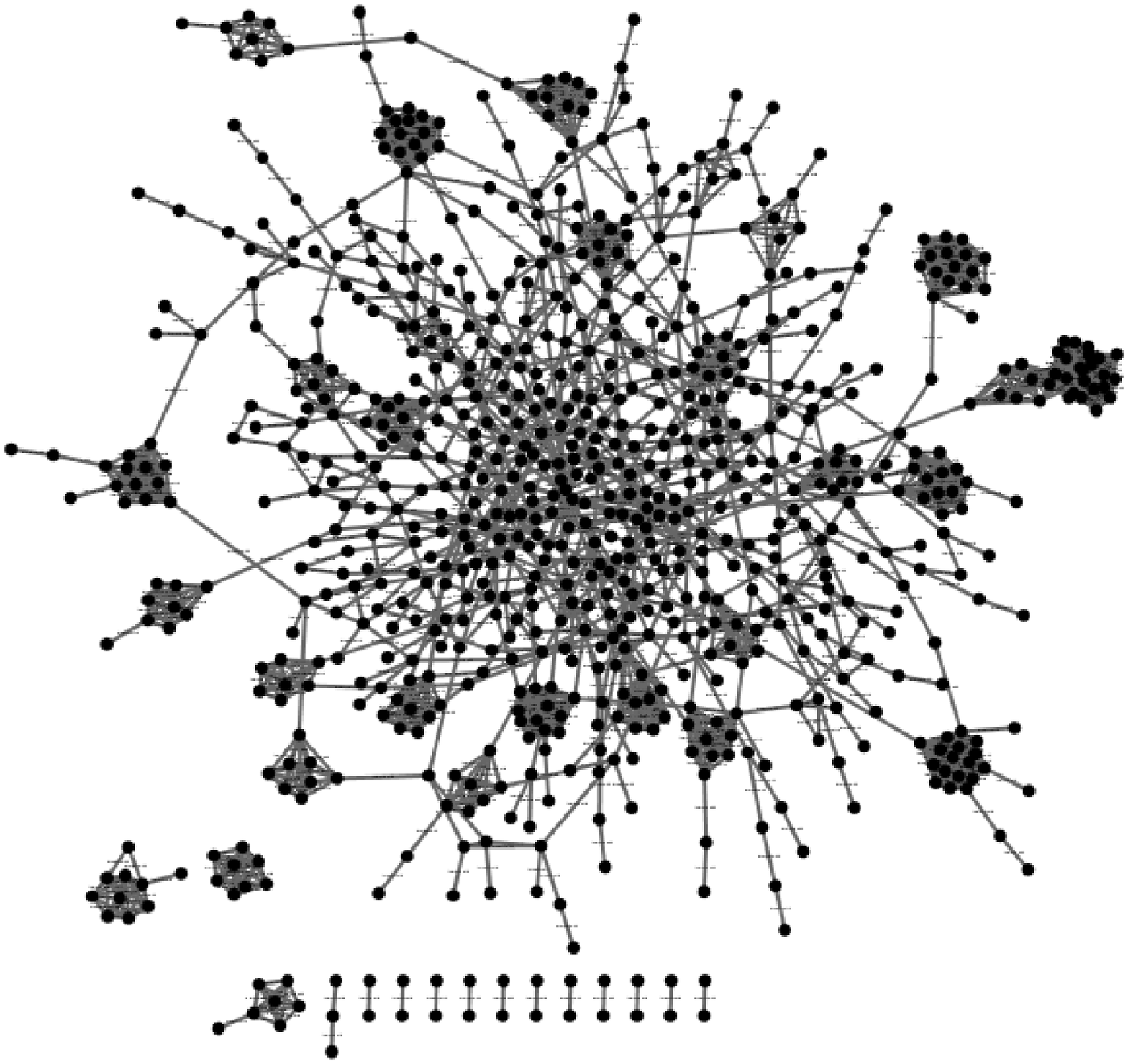}}
        \put(180,00){\includegraphics[width = 0.39\textwidth]{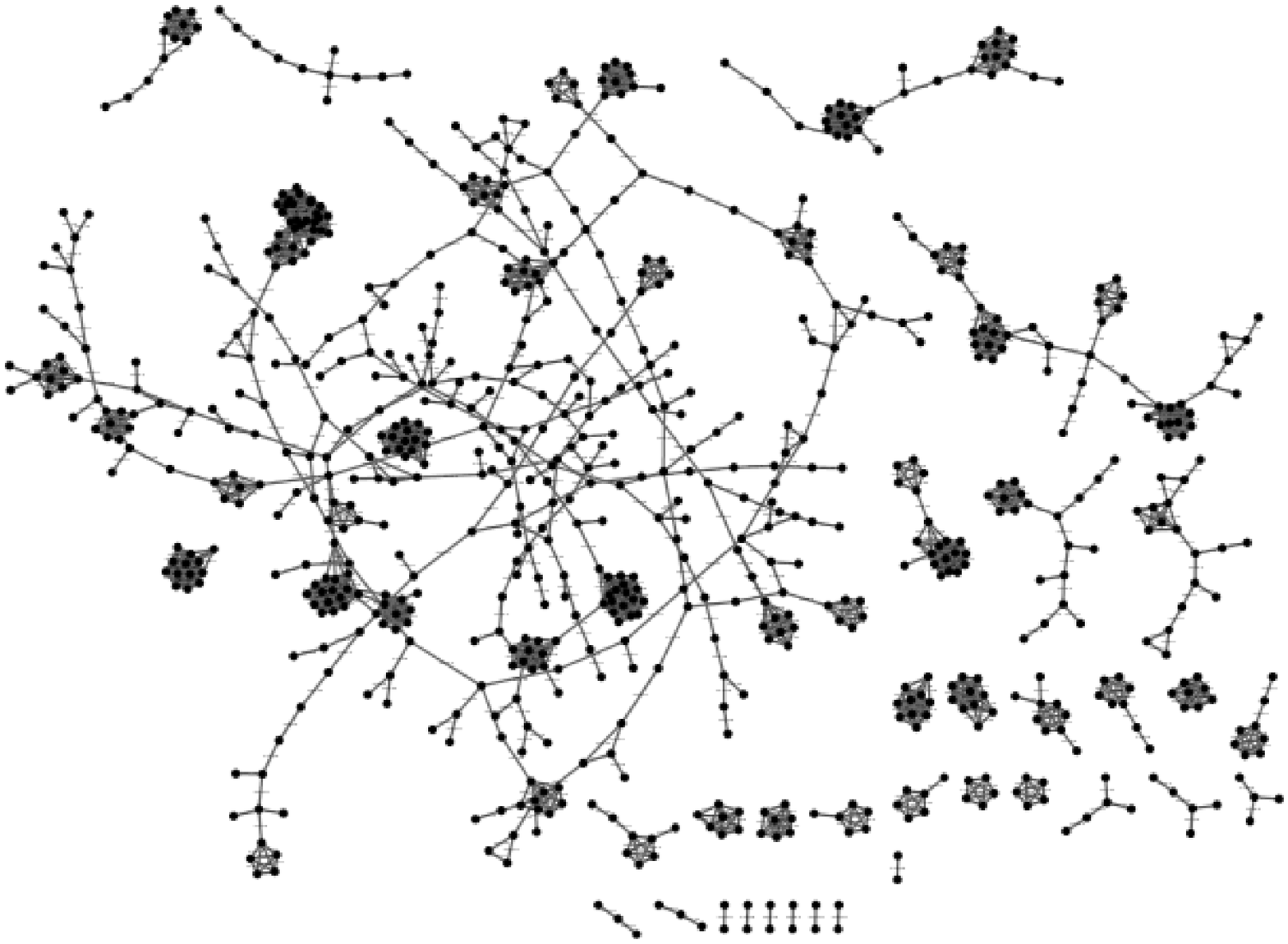}}
          \put(0,130) {$\alpha=0.5$:}  \put(0,118.5) {$m=24.3$}
           \put(180,130) {$\alpha=0.7$:}  \put(180,118.5) {$m=26.4$}
    \end{picture}
    \caption{
    Results of  numerical sampling simulations of the random graph ensemble (\ref{eq:ensemble}), with $N=1000$ and $p(k) = \frac{1}{4}\p{\frac{4}{5}}^k$.
    The four different images correspond to four different values of $\alpha$, with different loop densities $m(\alpha)$, as indicated.  Zero degree nodes are omitted.}
    \label{fig:graph-examples}
\end{figure}

In Figure \ref{fig:m-vs-alpha} we show the results of numerical sampling of graphs from (\ref{eq:ensemble}), using an appropriate MCMC process. The triangle density $m(\alpha)$ increases with $\alpha$, as expected. We observe distinct regimes of $\alpha$-values, as had already been observed for regular graphs in \cite{lopez2020imaginary}. Interestingly, to understand properly the nature of the different regimes of the ensemble it is necessary to also look at two other graph observables: the level of interaction between loops, measured with $r(\bA)$ as defined in (\ref{eq:rdef}), and the number $n(\bA)$ of connected components of the graph. We define their respective ensemble averages as $r(\alpha) = \expected{r(\bA)}$ and $n(\alpha) = \expected{n(\bA)}$.
The observed regimes are the following:

\begin{itemize}
    \item $\alpha\in[0,\alpha_1(N)]$: {\em Connected regime}
    \\[1mm]
    The loop density $m(\alpha)$ grows exponentially with $\alpha$, following
    \begin{align}
    \label{eq:malphat}
    m(\alpha) =\frac{1}{N} \p{\overline{k^2}/c -1 }^3\rme^{6\alpha}.
    \end{align} 
  Only the proportionality constant and the transition point $\alpha_1(N)$ depend on the degree distribution. This formula allows for an explicit calculation of $\alpha$, given a desired loop density, simply by inversion. For $\alpha=0$ it reproduces the rigorous result for the loop density for large graphs in \cite{bollobas1980probabilistic}. The degree of interaction between loops is as low as $r(\bA)\approx 3$ for large graphs. The number $n(\alpha)$ of components of the graph is the same as in the $\alpha =0$ case. It is relatively easy to obtain samples in this regime with the MCMC edge swap dynamics.
  
    \item $\alpha\in[\alpha_1(N),\alpha_2(N)]$: {\em Clustered regime}
    \\[1mm]
    Here the triangle density $m(\alpha)$ grows faster than (\ref{eq:malphat}). Depending on the chosen degree distribution, this growth may exhibit sudden jumps or  may be more smooth. The main difference with the previous regime is that loops start sharing edges. This follows from the observed drop of $r(\alpha)$. Nodes start to form clusters of similar degree. We call this the clustered regime, and $\alpha_1(N)$ the clustering transition point.
    
    \item $\alpha \in [\alpha_2(N),\infty)$: {\em Disconnected regime }
    \\[1mm]
    There is a drastic topological change associated with a second transition at $\alpha_2(N)$:  the graph breaks down into small disconnected cliques. Cliques of $k+1$ nodes  maximize the number of loops around a node of degree $k$, see Figures \ref{fig:graph-examples-2}  and \ref{fig:graph-examples}. Cliques associated with the maximum degree, with  $q+1$ nodes,  will appear first, followed by  those of the second largest degree, and so on. If, due to finite size effects, there are insufficient nodes to generate cliques, the graphs break down into small incomplete cliques. We call the transition at $\alpha_2(N)$ the \emph{shattering} transition, and this phase $\alpha>\alpha_2(N)$ the disconnected or shattered phase. The rest of the nodes, those unable due to degree constraints to form cliques, will continue to be connected and follow qualitatively similar regimes, but now for a new degree distribution that excludes the separated nodes.
\end{itemize}  

The transitions at $\alpha=\alpha_{1,2}(N)$ are not phase transitions in the conventional sense -- they depend on $N$, which is taken to be large but still finite -- so they are not marked by non-analyticities in the thermodynamic limit. The definitions given for $\alpha_1(N)$  and $\alpha_2(N)$ are instead of a descriptive nature, marking the $\alpha$-values where $\expected{r(\bA)}$ drops for the first time and where $\expected{n(\bA)}$ increases for the first time, respectively. As will become apparent in the next section, the system size $N$ affects severely the ensemble.

We make a distinction between bounded and unbounded degree distributions $p(k)$, since boundedness affects the way in which the ensemble behaves with increasing $N$, e.g. in the asymptotics of $\alpha_1(N)$ and $\alpha_2(N)$. For  large graphs with bounded degree distributions both transitions are close, $\alpha_1(N)\approx\alpha_2(N)$. There is a sudden appearance of disconnected cliques of $q+1$, nodes giving rise to a sharp jump in $m(\alpha)$.  
Strong numerical evidence and mathematical arguments support the proposition that $\alpha_1(N)$ and $
\alpha_2(N)$ both scale as $\Order{\log N}$. For graphs with unbounded $p(k)$, the maximum degree present in the graph will diverge slowly with $N$. Hence there are not many nodes of large degree to create cliques, and the structures created when the graphs shatter are less clear. The asymptotics of $\alpha_1(N)$ and $\alpha_2(N)$ should depend heavily on the tail of $p(k)$, as this tail governs the growth of the maximum degree with $N$. Nevertheless, in both cases the ensemble will end in a set of disconnected cliques, as this is the graph that maximizes the number of loops around each node. The difference between bounded and unbounded $p(k)$ can be  seen clearly when comparing Figure \ref{fig:graph-examples} and Figure \ref{fig:graph-examples-2}. For the graph with a bimodal degree distribution in Figure \ref{fig:graph-examples} the cliques appear immediately as the graph clusters, while for the one with an exponential distribution in Figure \ref{fig:graph-examples-2} one can see clusters appearing before the breaking down of the graph.

Expressions like (\ref{eq:freeEnergy}) are hard to evaluate analytically, especially for a finite $N$. The typical approach  of statistical mechanics would be to derive exact results in the limit $N\to\infty$, and then to show they are a good approximation for finite $N$. In contrast, here it is important \emph{not} to take the limit $N\to\infty$, but rather to work with asymptotically vanishing expressions  for the loop density, $m(\alpha) = \Order{N^{-\delta}}$. A clear example is that of the \emph{connected non interacting loopy regime}; here equation (\ref{eq:malphat}) shows correctly that $\lim_{N\to\infty}m(\alpha) =0$, but it is the way in which $m(\alpha)$  approaches $0$ that gives us formula (\ref{eq:malphat}), which is seen to be very accurate. One would normally rescale $\alpha$ with $N$ to avoid this effect, but it will become clear that in that case $m(\alpha_1(N))\to0$ for any proper scaling of $\alpha$ with $N$, meaning that the description of the first regime would vanish, which is not something we want. 

Regarding the sampling, we note that convergence from a given seed towards equilibration requires increasing numbers of edge swaps as $\alpha$ is increased. Only for values in the connected regime $\alpha\in[0,\alpha_1(N))$ will equilibration be fast enough to sample graphs in a reasonable amount of time on a personal computer. Close to the transitions there is a significant divergence of relaxation times. We conjecture that the main reason for this change is precisely the clustering of triangles: in order to break a clique one has to destroy many triangles, an event that becomes extremely unlikely during the dynamics for large graphs. Therefore we expect there to be an effective breaking of ergodicity when sampling with MCMC for $\alpha\!>\!\alpha_2(N)$ and large $N$.

For the above reasons, from the point of view of applied network science, working with the loopy ensemble (\ref{eq:ensemble}) has to be done carefully. Given a seed network, it is possible to randomize via edge swaps while retaining the value of the loop density, but there will be two problems. First, it could be that it takes a long time to sample correctly. Second,  it could be that samples generated with the same loop density have completely different topologies, according to their values of $r(\bA)$. The first problem is a matter of computing power and speed. The second problem is more tricky, and essentially unsolvable without modifying (\ref{eq:ensemble}). If the graph one wants to randomize has a value of $r(\bA)$ that deviates significantly from $\expected{r(\bA)}$, then all samples will be typically very different in structure, even though they share the same loop density.

\section{The connected regime \label{sec:analytic}}

We will now present an effective approximation for the generating function (\ref{eq:freeEnergy}). It is analogous to the one presented in \cite{lopez2020imaginary}, but generalized for an arbitrary degree distribution $p(k)$ with finite first and second moments.
We use a small $\alpha$ (or large $N$)  approximation to derive (\ref{eq:malphat}), using a known result about the distribution of triangles in the CM \cite{bollobas1980probabilistic}. It is found to give very good results, suggesting it could be exact asymptotically, at least for bounded degree distributions.  If we denote by $T(\bA)$ the number of triangles in $\bA$, we have (due to overcounting): 
\begin{align}
    \Tr(
    \bA^3) = 6 T(\bA).
\end{align}
We can therefore calculate the generating function (\ref{eq:freeEnergy}) as follows:
\begin{eqnarray}
\label{eq:phiCalculation}
    \phi(\alpha) &=& \frac{1}{N}\log \sum_{\bA}\rme^{6 \alpha T(\bA)}\prod_{i=1}^N \delta_{k_i,\sum_j A_{ij}} 
    \nonumber
    \\&=& \frac{1}{N}\log \sum_{T}
    \rme^{6\alpha T}P_N(T) + \frac{1}{N}\log \calN_{\bk}
\end{eqnarray}
Where we have introduced,
\begin{align}
    P_N(T) &= \frac{1}{\calN_\bk}\sum_{\bA} \delta_{T,T(\bA)}\prod_{i=1}^N \delta_{k_i,\sum_j A_{ij}}\\
    \calN_{\bk} &= \sum_{\bA}\prod_{i=1}^N \delta_{k_i , \sum_j A_{ij}}
\end{align}
Our approximation now consists in replacing $P_N(T)$ by the known asymptotic distribution of isolated triangles, that is triangles that do not share edges or nodes. The latter was computed rigorously in  \cite{bollobas1980probabilistic}:
\begin{eqnarray}
    P_N(T) &\approx & Poiss(T|\lambda_t) = \rme^{-\lambda_t}\frac{(\lambda)^T}{T!}\\
    \lambda & =& \frac{1}{6}\frac{\sum_{i=1}^N k_i (k_i-1)}{\sum_{i=1}^N k_i} = \frac{1}{6}\p{\frac{\overline{k^2}}{c}-1}
\end{eqnarray}
This then leads us to the the following approximation for (\ref{eq:phiCalculation}) and $m(\alpha)$:
\begin{eqnarray}
    \phi(\alpha) &\approx & \frac{1}{N} \lambda_T \p{\rme^{6\alpha} - 1} + \frac{1}{N}\log \calN_\bk \\
    m(\alpha) & \approx & \frac{1}{N}6\lambda_T \rme^{6\alpha} = \frac{1}{N} \Big(\overline{k^2}/c-1\Big)\rme^{6\alpha}
\end{eqnarray}
This formula has a simple interpretation. At $\alpha = 0$ it correctly predicts the expected number of triangles in a CM, where one pictures these triangles to be very far away from each other. When $\alpha>0$ this number of triangles is multiplied by $\rme^{6\alpha}$, giving another finite but larger amount of triangles when $N\to \infty$. In this scenario we would view these triangles to be simply further and further apart as the system size grows. This picture will be revisited in the next section.

\begin{figure}[t]
    \centering
    \begin{picture}(360,145)
        \put(0,0){\includegraphics[width = 0.52\textwidth]{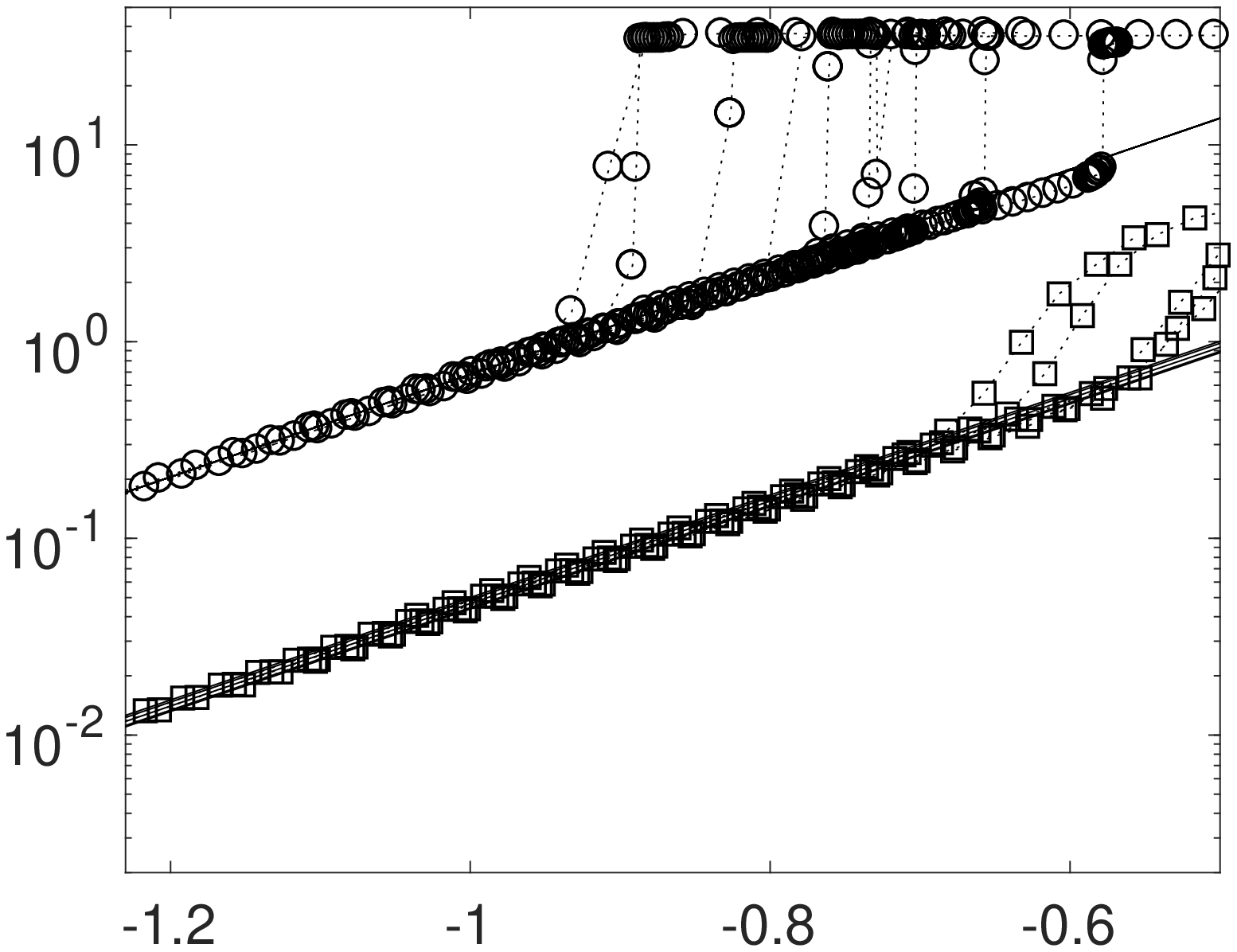}}
        \put(180,0){\includegraphics[width = 0.52\textwidth]{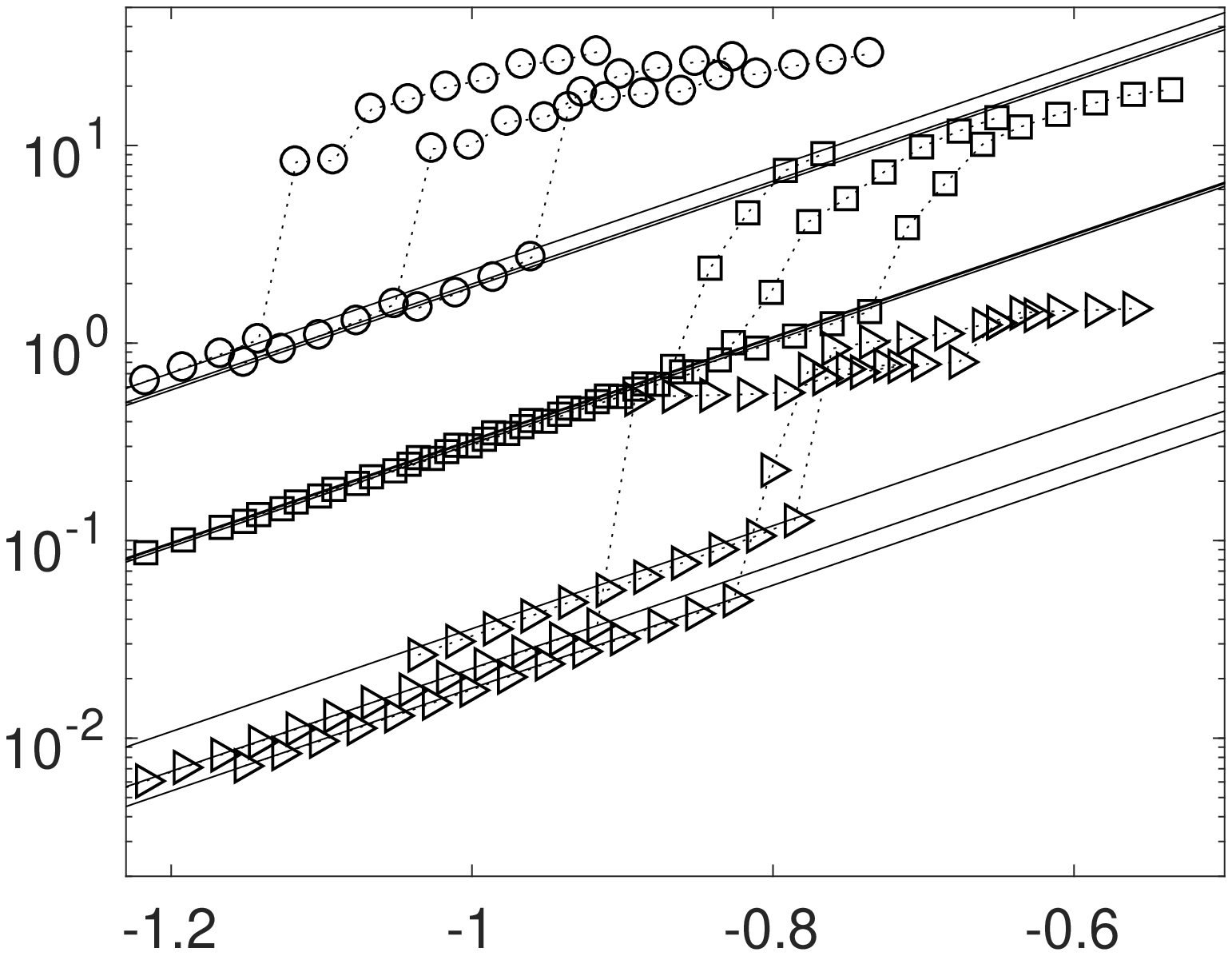}}
        \put(90,0){$\Tilde\alpha$}
        \put(270,0){$\Tilde\alpha$}
        \put(0,80){$m$}
    \end{picture}
    \caption{The loop density $m$ as measured in numerical MCMC simulations of the ensemble (\ref{eq:ensemble}), plotted against the rescaled control parameter $\Tilde{\alpha} = \alpha - \frac{1}{6}\log N$. Left panel: $p(k) =\frac{1}{2}\delta_{k3}+\frac{1}{2}\delta_{k9}$ (circles, for system sizes $N = 100,200,300,400,500,750,1000,2000,4000$, from right to left), and $p(k) =\frac{1}{5}\sum_{j=1}^5 \delta_{kj}$  (squares, for system sizes $N = 500,750,1000,2000$, from right to left). Right panel: $p(k)=exp(k|5)$ (circles),  $p(k) = Poiss(k,5)$ (squares), and $p(k) =PL(k)$ (triangles), all for system sizes $N=500,1000,2000$. See Table   \ref{tab:degreeDistributions} for the relevant definitions.  Error bars were omitted for clarity. 
    The solid lines correspond to the corresponding theoretical prediction (\ref{eq:m-alpha-tilde}).}
    \label{fig:trianglesLowDensity}
\end{figure}

We have tested the above approximation extensively with numerical simulations. We generated samples from (\ref{eq:ensemble}) for many different degree distributions, shown in Table \ref{tab:degreeDistributions}. The results are shown in Figure \ref{fig:trianglesLowDensity}, where we have plotted the results for systems of multiple sizes $N\sim 100-4000$. In order to have a better visualization, we plotted the loop densities  against a rescaled parameter $\Tilde{\alpha}$, defined via $\alpha=\Tilde{\alpha} + \frac{1}{6}\log N$,
\begin{align}
\label{eq:m-alpha-tilde}
    m\big(\Tilde{\alpha} +\frac{1}{6}\log N\big) \approx \Big(\frac{\overline{k^2}}{c}-1\Big)^3 \rme^{6\Tilde{\alpha}} \hspace{15mm}\textrm{ for }\Tilde{\alpha}\le\Tilde{\alpha}_1(N).
\end{align}

For this regime we used waiting times of $2\cdot 10^4$ AESPL (Attempted Edge Swaps Per Link), and subsequently recorded $20$ samples spaced by $2\cdot 10^3$ AESPL.  To show the accuracy of the theory with a modest number of samples, we plot the average of the loop density over the full time series of loop densities between samples. We do this to reduce noise, and because our theory refers to the average (\ref{eq:malpha}), not to graph instances,  since there is no self averaging at finite sizes. For graphs larger then $500$ nodes, error bars are  of the order of magnitude of the markers. For smaller graphs the error bars can be appreciated on the right panel of Figure \ref{fig:transition}. In the remaining loop density plots the error bars were omitted, in order to avoid cluttering of figures. 
Note that the scaling in (\ref{eq:m-alpha-tilde}) collapses all curves of the same degree distribution, up to a certain value $\Tilde{\alpha_1}(N)$. As we will show in the next section, the loop density at the transition vanishes as $N\to\infty$, $m(\Tilde{\alpha}_1(N))\to 0$. This can be clearly seen in Figure \ref{fig:trianglesLowDensity}.

\begin{table}

    \caption{Different degree distributions used for numerical experiments.}
   \begin{tabular}{@{}llll}
    \br
    type    & name & formula $p(k)$ & parameter values \\
    \mr
    unbounded    &  exponential  & $ exp(k|c)=\p{\frac{c}{c+1}}^k\frac{1}{c+1} $ & $c = 3,4,5,10 $ \\[0.5mm]
    unbounded     & Poissonian & $Poiss(k|c) =  \rme^{-c}\frac{c^k}{k!}$ & $c = 3,4,5,10$\\[2mm]
    unbounded & power law & $PL(k)=A k^{-\gamma} $ ~ $k\ge 2$ & $\gamma = 4$ ($\overline{k} \approx 2.5$)\\[1mm]
    \mr
    bounded & bimodal &   $bim(k|3,q)=\frac{1}{2}(\delta_{k,3} + \delta_{k,q}) $ & $q = 5,7,9$\\[1mm]
    bounded & uniform & $u(k) =  \frac{1}{5}\sum_{j=1}^5 \delta_{k,j}$ & - \\[1mm]
    bounded & non uniform & $v(k) =  \sum_{j=1}^5 w_j \delta_{k,j}$ & $\boldsymbol{w} = (\frac{1}{10},\frac{2}{10},\frac{3}{10},\frac{3}{10},\frac{1}{10})$\\[1mm]
    \br
    \end{tabular}
    \label{tab:degreeDistributions}
\end{table}

The accuracy of (\ref{eq:malphat}) suggests that it could be the exact asymptotic result when $N \to \infty$. This would imply that a bias of the form (\ref{eq:ensemble}) with $\alpha = \Order{1}$ only modifies the number of expected triangles in large graphs by an $\Order{1}$ amount, implying that the loop density will still vanish asymptotically. To achieve a nonvanishing loop density in the asymptotic limit, a different scaling of $\alpha$ should be introduced, as was done in \cite{lopez2018exactly} for 2-regular graphs, i.e. for $p(k) = \delta_{k,2}$. However, as will be discussed in the next section, for general degree distributions the effect of scaling $\alpha$ with $N$ is much more complicated than in the 2-regular case.

\section{The clustered and disconnected regimes}

\subsection{General results}

\begin{figure}
    \centering
    \begin{picture}(360,273)
    \put(50,179){\includegraphics[trim = 0 60 0 0,clip,width=0.8\textwidth]{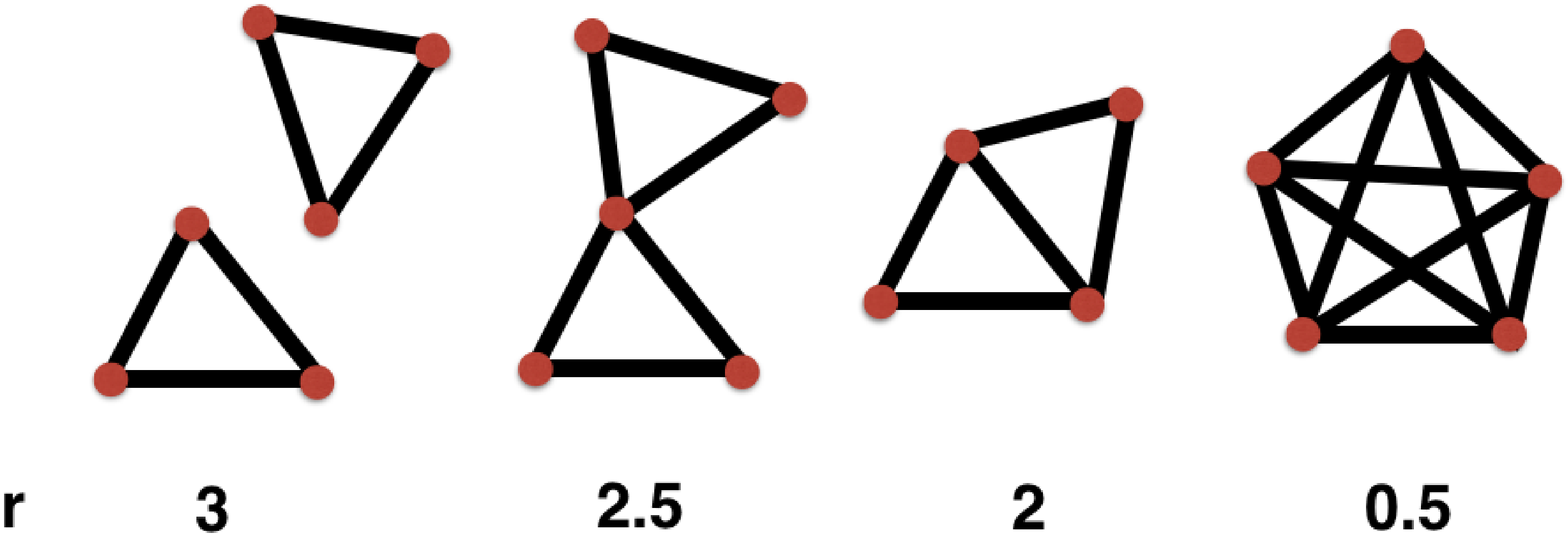}}
    \put(10,10){\includegraphics[width=0.5\textwidth]{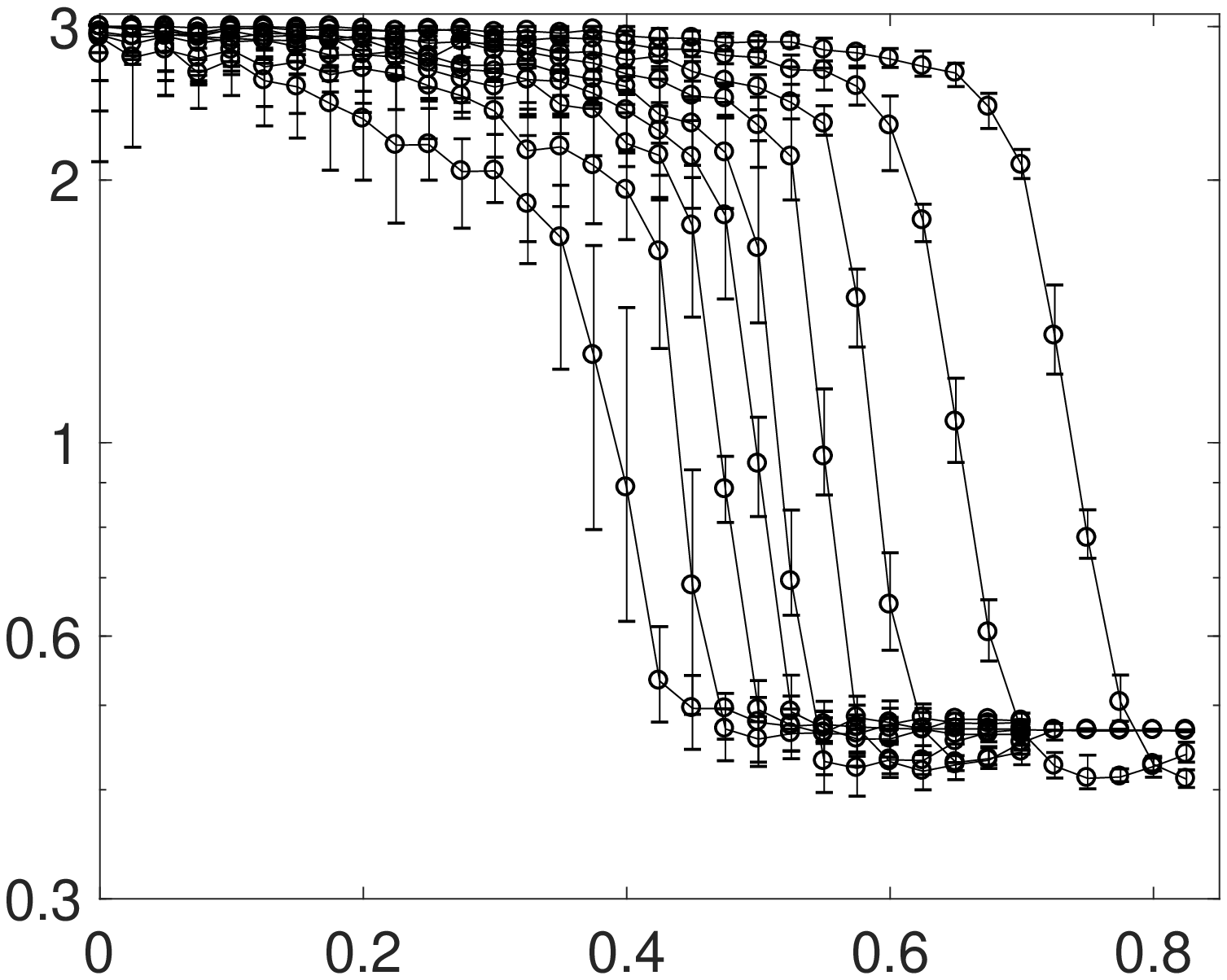}}
    \put(180,10){\includegraphics[width=0.5\textwidth]{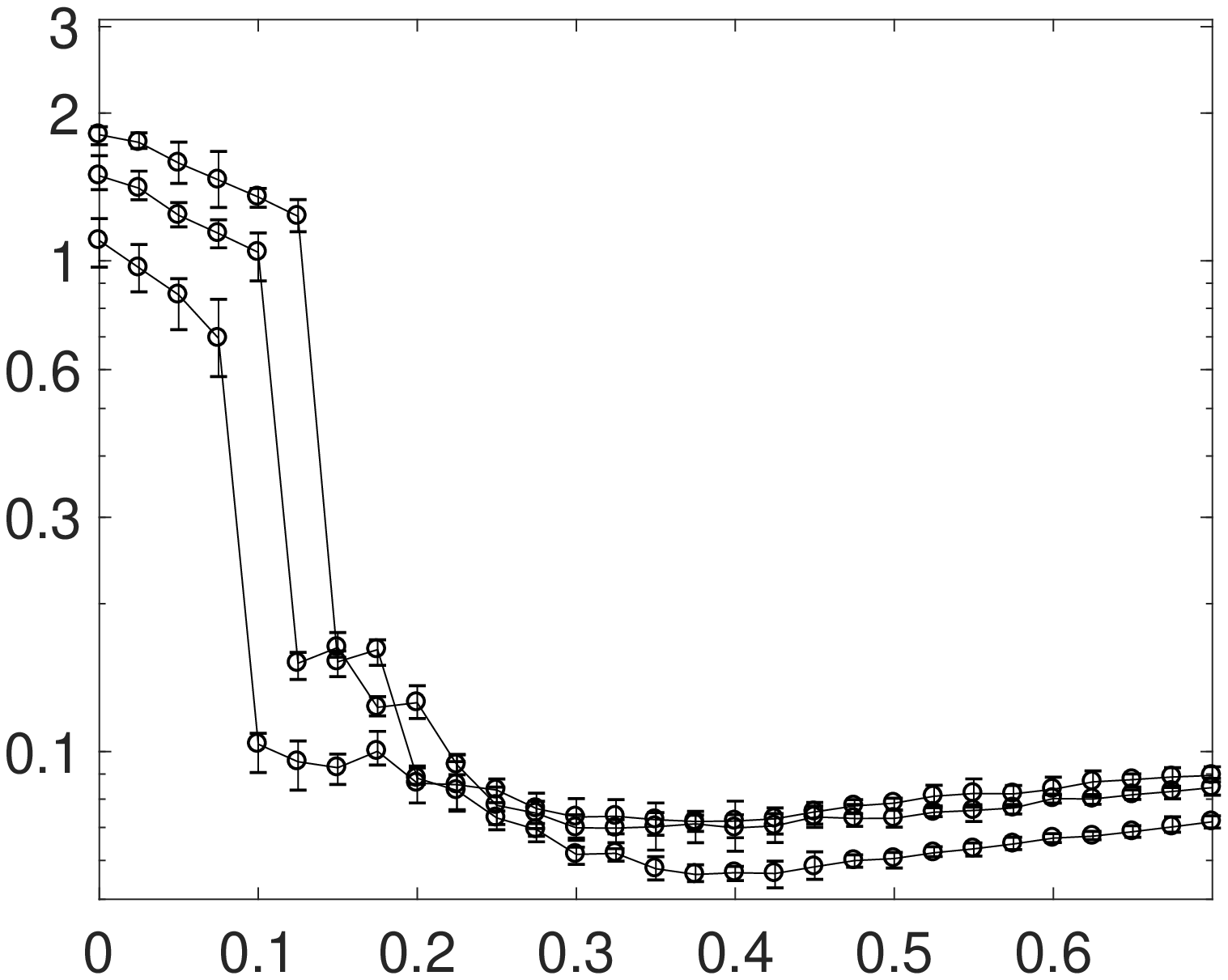}}
    \put(90,3){$\alpha$}
    \put(270,3){$\alpha$}
    \put(0,80){$r(\alpha)$}
    \put(77,175){$r(\bA) = 3$}
    \put(152,175){$r(\bA)=2.5$}
    \put(222,175){$r(\bA)=2$}
    \put(292,175){$r(\bA)=0.5$}
    \end{picture}
    \caption{ Top row: examples of small graphs and their corresponding values of $r(\bA)$.
    Bottom row: plots of $r(\alpha)$ as measured in simulations, shown versus $\alpha$, with standard deviations shown as error bars. Left:  $p(k) = \frac{1}{2}\delta_{k3}+\frac{1}{2}\delta_{5k}$, with graph sizes $N = 100,200,300,400,500,750,1000,2000, 4000$ (from left to right).
    Right: $p(k) =\rme^{-5} 5^k/k!$, with graph sizes $N = 500,1000,2000$ (from left to right). }
    \label{fig:rExamples}
\end{figure}

We next investigate the behaviour of the ensemble beyond the clustering transitions, i.e. for $\alpha >\alpha_1(N)$, where (\ref{eq:malphat}) no longer reproduces the correct loop density.
For the 2-regular case, the only loopy structure that can exist inside a graph is an isolated cycle, therefore it is possible for (\ref{eq:m-alpha-tilde}) to be exact  asymptotically. For other degree distributions, many other loop structures can appear in a graph. As we will show, it seems that structures with strongly interacting triangles dominate entropically. Therefore the statistics of different local structures needs to be taken into account, making (\ref{eq:malphat}) insufficient to describe the ensemble for all values of $\alpha$.

In the regime  $\alpha<\alpha_1(N)$, the desired loop density is achieved by creating further triangles that are independent and far from each other, without sharing nodes. For $\alpha>\alpha_1(N)$, in contrast, the desired loop density is achieved by creating triangles that share as many edges as possible. This qualitative change  appears  to be purely entropic, since the latter regime appears for all loop densities as long as the system is large enough, that is even for very small values of $m$. 
Put differently, the transition at $\alpha_1(N)$ does \emph{not} happen because there are too many triangles which need  to share nodes due to of lack of space in the graph, as one might guess initially. The transition happens because for a given loop density the number of graphs one can create by `putting triangles aside' in small clusters is larger than the number of graphs one can create by embedding them in the graph in a non-interacting way.  While we cannot prove this assertion rigorously,  extensive numerical experiments support this claim. 

We measured the interaction between loops in samples of (\ref{eq:ensemble}) using the observable $r(\bA)$ defined in (\ref{eq:rdef}). The empirical value $r(\alpha) = \expected{r(\bA)}$ was measured in all the numerical experiments listed in table \ref{tab:degreeDistributions}. For values $\alpha>\alpha_1(N)$ we increased the number of AESPL by a factor ten, giving waiting times of  $2\!\cdot\! 10^5$ AESPL and inter-sample intervals of $2\!\cdot\! 10^4$ AESPL. In all experiments we observed the same behaviour as shown for the two cases in Figure \ref{fig:rExamples}. An initial phase of $\expected{r(\bA)}\approx 3$, indicating non-interacting loops, is followed by a sudden drop to $\expected{r(\bA)}  = r_{min}(N)<1$, indicating interacting loops. At the value of $\alpha$ marking this sudden drop, which we defined to be $\alpha_1(N)$, the graph has become clustered in order to achieve the desired loop density. This $\alpha$ value coincides precisely with the point where formula (\ref{eq:malphat}) stops working, as can be seen in Figure \ref{fig:m-vs-alpha}.

When increasing the system size $N$, it is clear that the initial parts of the curves tend to flatten to  plateaux at the level $r = 3$. This is consistent with the fact that equation (\ref{eq:malphat}), which accurately describes the loop density in this regime, was derived assuming an underlying Poissonian distribution of triangles; the latter  assumes, in turn, that the triangles are non-interacting \cite{bollobas1980probabilistic}.
\vspace*{3mm}

The remaining question is how the two values $\alpha_1(N)$ and $r_{min}(N)$ depend on $N$.
For $r_{min}(N)$ the following possibilities must be considered:
\begin{enumerate}
    \item $\lim_{N\to\infty}r_{min}(N) = r^* > 0$
    \item $\lim_{N\to\infty}r_{min}(N) =  0$ 
\end{enumerate}
Given that for a finite graph $r(\bA)$ is always bounded from below by $r = 6/(q^2 - q)$, the second option is only a possibility for unbounded graphs.
For $\alpha_1(N)$ we have the following possibilities, with their different physical implications:
\begin{enumerate}
    \item $\lim_{N\to\infty}\alpha_1(N) = \infty$, asymptotically the loop density vanishes for all values of $\alpha$.
    \item $\lim_{N\to\infty}\alpha_1(N) = \alpha^* >0$, there is a first order phase transition at $\alpha^*$.
    \item $\lim_{N\to\infty}\alpha_1(N) = 0$, all $\alpha$ values have a finite density loop density  $m(\alpha)>0$.
\end{enumerate}
We made the distinction between bounded and unbounded distributions precisely because we believe that the behaviour of the ensemble for  these distribution families  might not be the same. As can already be seen in the bound $r(\bA)\ge 6/(q^2-q)$, if $q$ is growing with $N$, then $r$ can approach the value $0$ arbitrarily closely, contrary to the bounded case. This can also be appreciated in Figure \ref{fig:rExamples}, for the exponentially distributed degree distribution  $\expected{r(\bA)}$ appears to reach a lower value for larger $N$. 

For the case of bounded distributions, the maximum degree $q$ asymptotically provides sufficiently many nodes to create cliques that will achieve the desired loop density, see for example Figure \ref{fig:graph-examples}. If the desired loop density is higher, then this density will be realized via cliques of the next highest degree $k<q$, in descending order. For unbounded degree distributions, this picture changes.  Here one cannot guarantee the abundance of such cliques, therefore the observed topology seems to remain connected for larger values of $\alpha$, in what we have called the clustered regime. 

\subsection{Results for bounded degree distributions}

\begin{figure}
    \centering
    \begin{picture}(360,150)
        \put(0,10){\includegraphics[width = 0.5\textwidth]{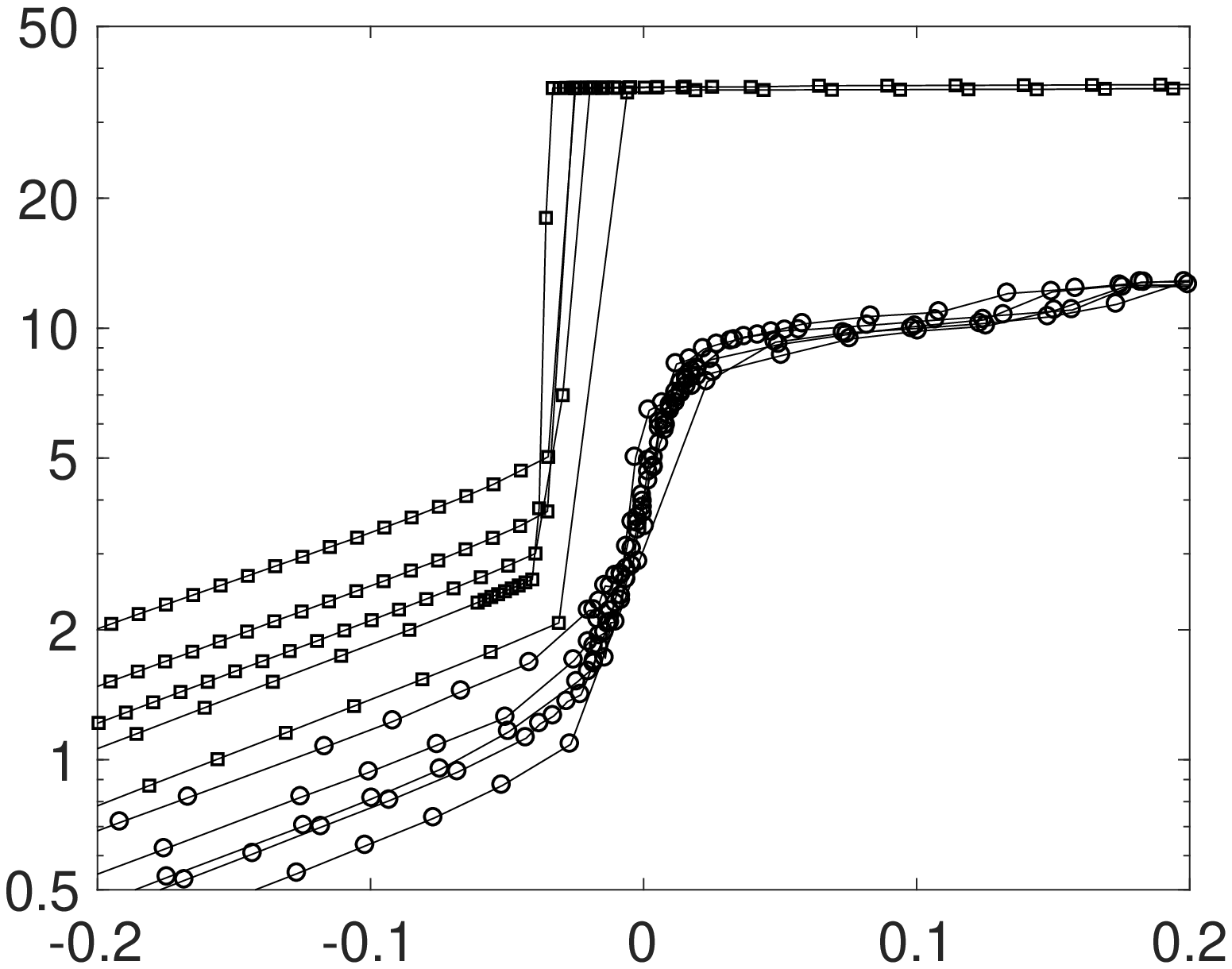}}
        \put(180,10){\includegraphics[width = 0.5\textwidth]{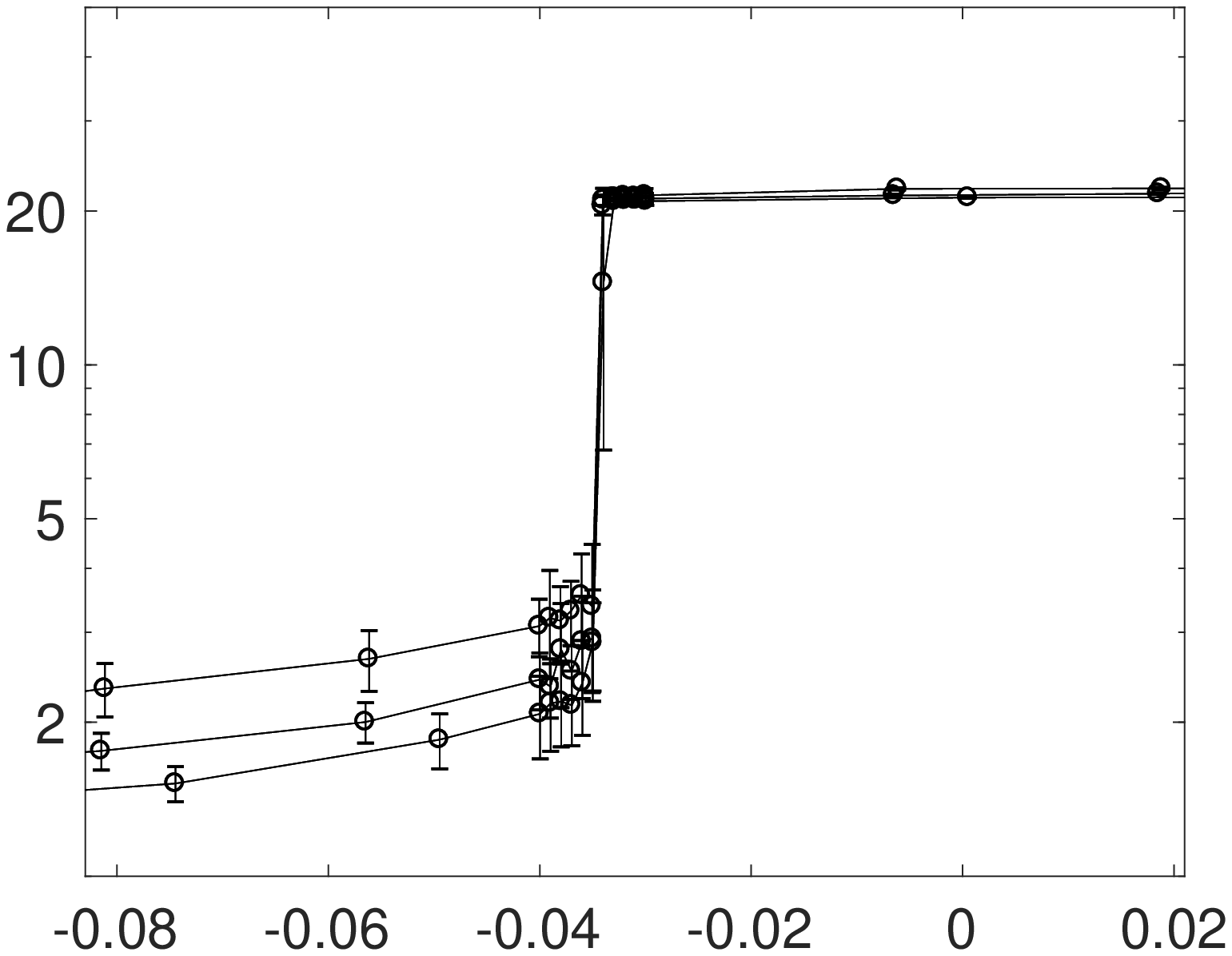}}
        \put(93,0){$\gamma$}
        \put(270,0){$\gamma$}
        \put(0,80){$m$}
    \end{picture}
    \caption{Plots of $m$ against rescaled variable $\gamma = \alpha - [2(q+1)]^{-1}\log N$, showing the collapse of the second (shattering) transition point for different system sizes, predicted by (\ref{eq:m-gamma}). Left: $p(k) =\frac{1}{2}\delta_{k3}+\frac{1}{2}\delta_{k5}$ (circles), and  $p(k) = \frac{1}{2}\delta_{k3}+\frac{1}{2}\delta_{k9}$  (squares). System sizes were $N = 200,300,400,500,750$, from bottom to top. Error bars are omitted to reduce cluttering. Right: close-up in the neighbourhood of the shattering transition, for $p(k) = \frac{1}{2}\delta_{k3}+\frac{1}{2}\delta_{k7}$, for system  sizes $N=200,300,400$ (from bottom to top). Here the error bars correspond to average plus/minus one standard deviation.}
    \label{fig:transition}
\end{figure}

In this subsection we develop a further theoretical  description of our graph ensemble   for the case of bounded degree distributions. As mentioned before, numerical simulations suggest the need to include the statistics of the cliques formed by nodes of maximum degree. We denote by $K_q(\bA)$ the number of fully connected cliques of $q\!+\!1$ nodes, and by $T(\bA)$ the number of triangles that are not in cliques of degree $q$. We can then decompose the total number of 3-loops in the following way:
 \begin{align}
    \Tr(\bA^3) = 6 T(\bA) + (q+1)q(q-1)K_q(\bA).
\end{align}
With this decomposition we can write the partition function as
\begin{eqnarray}
    \phi(\alpha)&=& \frac{1}{N} \log \sum_{\bA} \rme^{6\alpha T(\bA) + \alpha (q+1)q(q-1)K_q(\bA)}\prod_{i = 1}^N \delta_{k_i,\sum_j A_{ij}}\nonumber\\
     &= &\frac{1}{N}\log \sum_{T,K} P_N(T,K) \rme^{6\alpha T}\rme^{q(q^2-1)\alpha K } + \frac{1}{N}\log \calN_{\bk}
\end{eqnarray}
where we introduced $  \calN_{\bk} =  \sum_\bA \prod_{i\leq N} \delta_{k_i,\sum_{j}A_{ij}}$, and  the joint distribution of triangles and cliques for the unbiased CM, 
\begin{eqnarray}
    P_N(T,K) &=&  \frac{1}{\calN_{\bk}}\sum_\bA  \delta_{T,T(\bA)}\delta_{K,K_q(\bA)} \prod_{i=1}^N \delta_{k_i,\sum_{j}A_{ij}}.
\end{eqnarray}
Our main approximation consists in assuming that asymptotically the random variables $T$ and $K$ become independent, each described by Poisson distribution. This means that we again assume the main contribution of triangles for $T(\bA)$ to come from isolated triangles. Since isolated triangles and cliques are almost independent, and are rare events in the CM, one could argue that according to the \emph{Poisson Paradigm}  in \cite{alon2004probabilistic}, they should both be Poissonian random variables. For a similar argument regarding loops of different lengths see \cite{wormald1999models}. Thus we put
\begin{align}
\label{eq:poissonian}
    P_{N}(T,K) \sim Poiss(T|\lambda_t)Poiss(K|\lambda_{K_q}(N))
\end{align}
We can then immediately proceed to calculate the partition function,
\begin{eqnarray}
    \phi(\alpha) &\approx &  \frac{\lambda_T}{N}\p{\rme^{6\alpha} - 1} +  \frac{\lambda_{K_q}(N)}{N}\p{\rme^{q(q^2-1)\alpha} - 1} + \frac{1}{N}\log \calN_{\bk},
\end{eqnarray}
which leads to the following expression for the loop density,
\begin{align}
      m(\alpha) \approx & \frac{6\lambda_T}{N}\rme^{6\alpha} + \frac{q(q^2-1)}{N}\lambda_{K_q}(N) \rme^{q(q^2-1)\alpha}.
\end{align}
Contrary to the regular case discussed in \cite{lopez2020imaginary}, there is for an arbitrary $p(k)$ no established rigorous result for the expected number $\lambda_K(N)$ of cliques. Nevertheless, there is a good idea of what its scaling with $N$ should be \cite{bollobas2001random}. The expected number of isomorphisms of a given strictly balanced graph $H$ (see \cite{bollobas2001random} for definition) is expected to be $\Order{N^{v(H)-e(H)}}$, where $e(H)$ and $v(H)$ are the number of edges and nodes of $H$ respectively. In the case of a clique of $q+1$ nodes these numbers are, $e(K_{q}) = \frac{1}{2}q(q+1)$ and $v(K_{q}) = q+1$. Therefore,
\begin{align}
    \lambda_{K}(N) = \Order{\frac{1}{N^{\frac{1}{2}q(q-1)-1}}}\sim\frac{c_q}{N^{\frac{1}{2}q(q-1)-1} q(q^2-1)}.
\end{align}
We have included the factor $q(q^2-1)$ in the denominator for convenience. With this expression we obtain the following result for small values $\alpha$
\begin{align}
    \label{eq:malphaK}
    m(\alpha) \approx \frac{1}{N}\p{\overline{k^2}/c-1}^3\rme^{6\alpha} + \frac{c_q}{N^{\frac{1}{2}q(q-1)}}\rme^{\alpha q(q^2 -1)}
\end{align}
The first term corresponds to the contribution from isolated triangles at low density, to be denoted by $m_t(\alpha)$. The second term represents triangles  in the previously described cliques, we denote is as $m_K(\alpha)$. The latter is bounded since the number of cliques of $q +1$ nodes is bounded. This then gives
\begin{align}
    m_K(\alpha) \approx \left\lbrace\begin{array}{cc}
    N^{-\frac{1}{2}q(q-1)}c_q \rme^{q(q^2-1)\alpha}     &\textrm{ if } \alpha\le\alpha_2(N) \\[1mm]
    p(q) q (q-1)     & \textrm{ if } \alpha\ge\alpha_2(N)
    \end{array}\right.
\end{align}
It is convenient to define the shattering transition as the point where all the cliques of degree $q$ have appeared. This automatically gives an estimate of how $\alpha_2(N)$ behaves with $N$. Here we can see that $ \alpha_2(N) $ diverges logarithmically with $N$:
\begin{align}
\label{eq:alpha2}
    \alpha_2(N) = \frac{1}{2(q+1)}\log N + \frac{1}{q(q^2-1)}\log\s{\frac{p(q)q(q-1)}{c_q}}
\end{align}
This result depends on the degree distribution $p(k)$ explicitly through $q$ and $p(q)$, but also implicitly through $c_q$. Since we do not generally know $c_q$, we can not test the accuracy of the above prediction directly. Only for regular graphs $c_q$ is available, leading to accurate predictions for $\alpha_2(N)$  \cite{lopez2018exactly}. However, alternative tests are possible. Equations (\ref{eq:alpha2}) predicts a collapse of the various $  \alpha_2(N) $ curves under the following change of variable, $\alpha = \gamma + \frac{1}{2(q+1)}\log N$,
\begin{align}
\label{eq:m-gamma}
    m \p{\gamma  + \frac{1}{2(q+1)}\log N}  \approx \left\lbrace \begin{array}{cc}
        N^{-\frac{q-2}{q+1}} \rme^{6\gamma}& \textrm{ for }  \gamma\le\gamma_1(N) \\[1mm]
        c_q \rme^{q(q^2-1)\gamma} &\textrm{ for } \gamma_1(N)\le \gamma \le \gamma_2(N)
    \end{array}\right.   
\end{align}
Even though it is hard to sample graphs very precisely in the clustering regime, given that the waiting time of the MCMC algorithm is very large, overall the transition points of the curves do collapse nicely, as can be seen in Figure \ref{fig:transition}. We stress that close the transition waiting times were so long that points on the steep part of the left panel on Figure \ref{fig:transition} were probably not equilibrated for system sizes $N\ge 1000$. For this reason we show in the right panel  that for system sizes $N=200,300,400$ we do see an almost perfect collapse of the transitions points of the curves. For these small sizes was it possible to have confidence in the equilibration of the MCMC algorithm so close to the transition. The prefactor slope of $\frac{1}{2(q+1)}$ for  the term proportional to $\log N$ in $\alpha_2(N)$ in (\ref{eq:alpha2}) was also tested. Results are presented in Table \ref{tab:comparison}. We find a very good agreement for the bimodal distributions. For distributions $u(k)$ and $v(k)$ the prediction is close enough to the predicted value $0.8\overline{3}$, but the observed value of $0.10(1)$ in both cases is actually closer to what we would observe with $q=4$. This is consistent with the fact that, for these particular distributions, both degrees have a similar density and $k=4$ is more abundant in the case of $v(k)$.

\begin{figure}[t]
    \centering
    \begin{picture}(330,300)
        \put(10,170){\includegraphics[width = 0.46\textwidth]{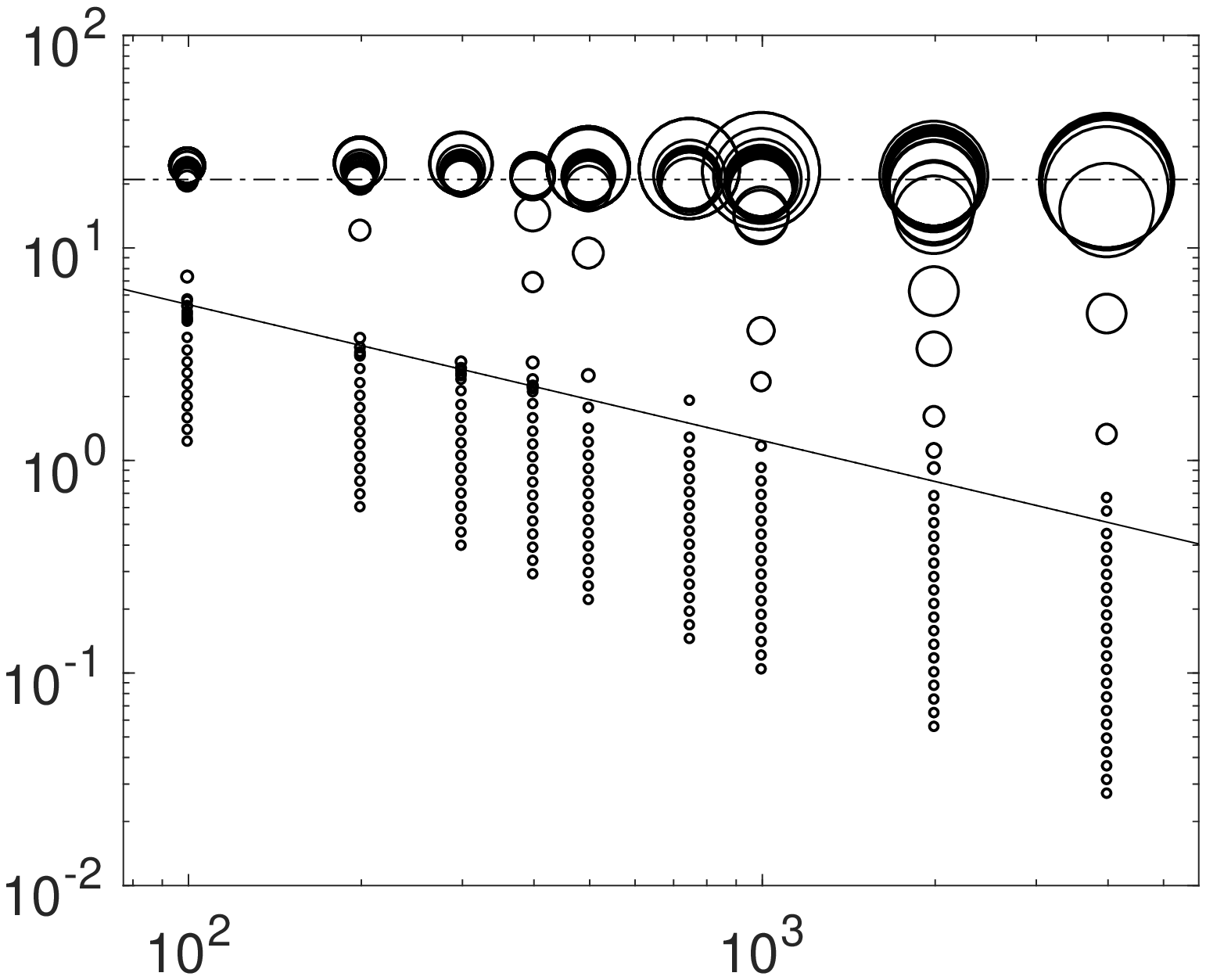}}
         \put(180,170){\includegraphics[width = 0.46\textwidth]{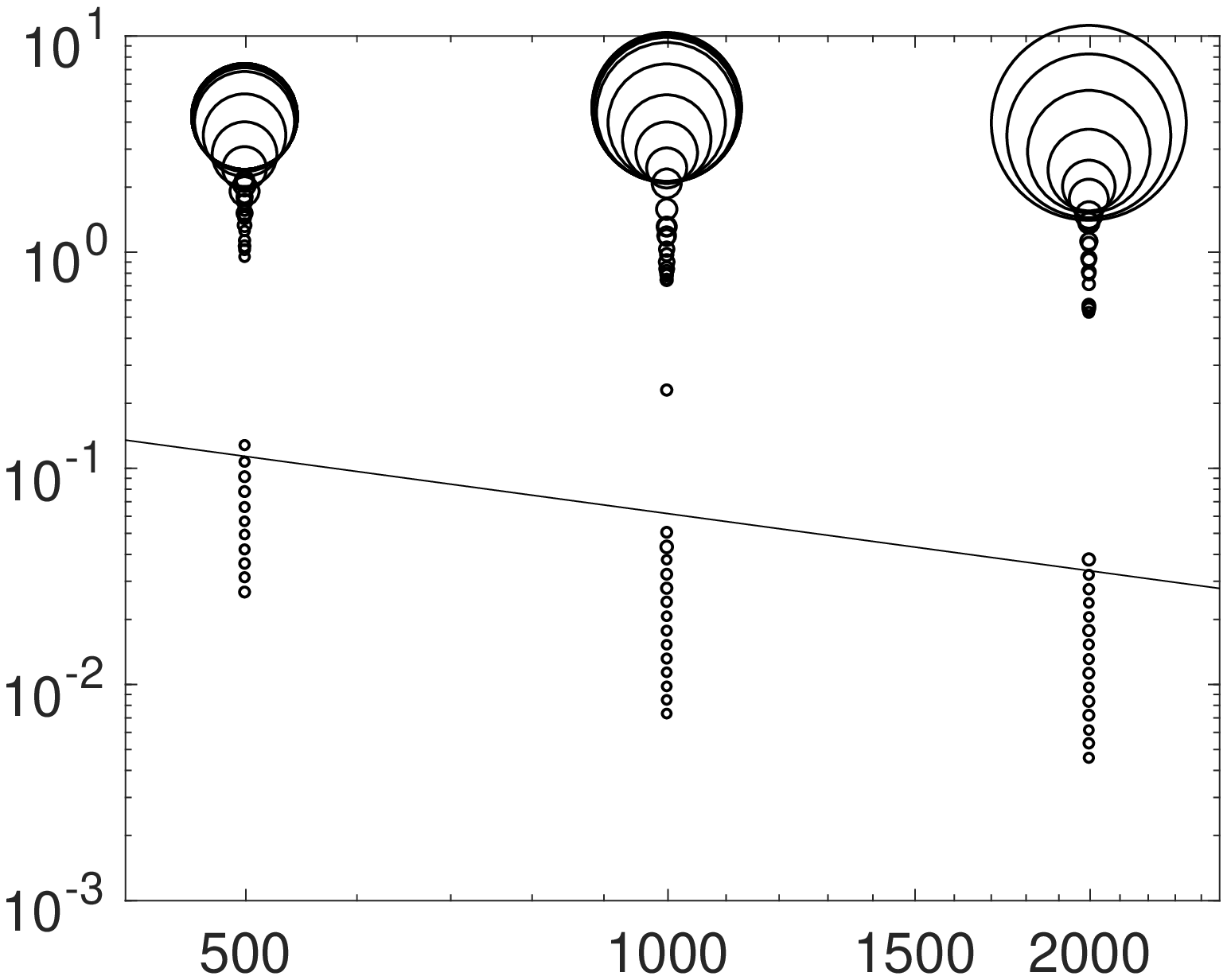}}
        \put(203,187){\includegraphics[width = 0.13\textwidth]{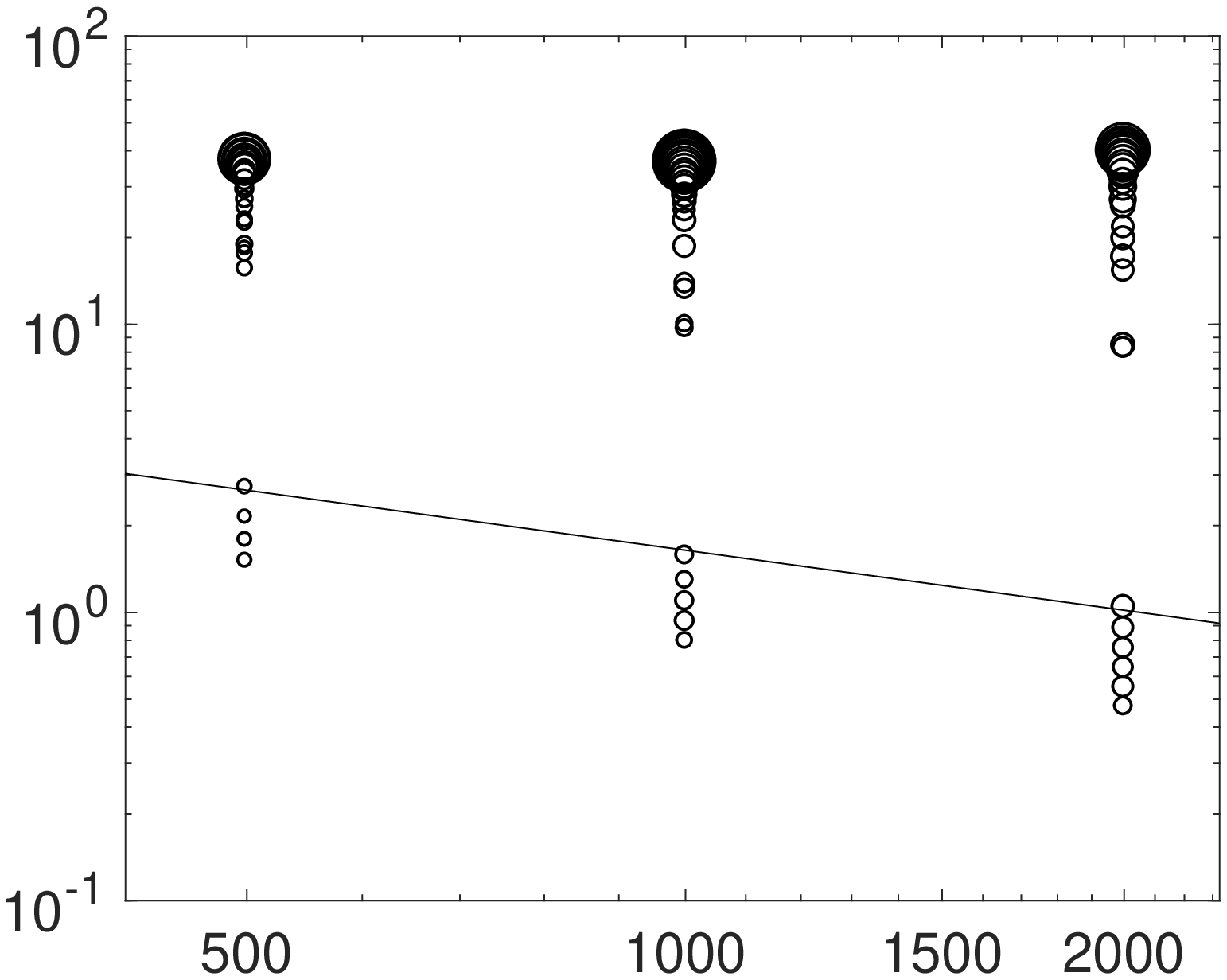}}
        \put(10,25){\includegraphics[width = 0.46\textwidth]{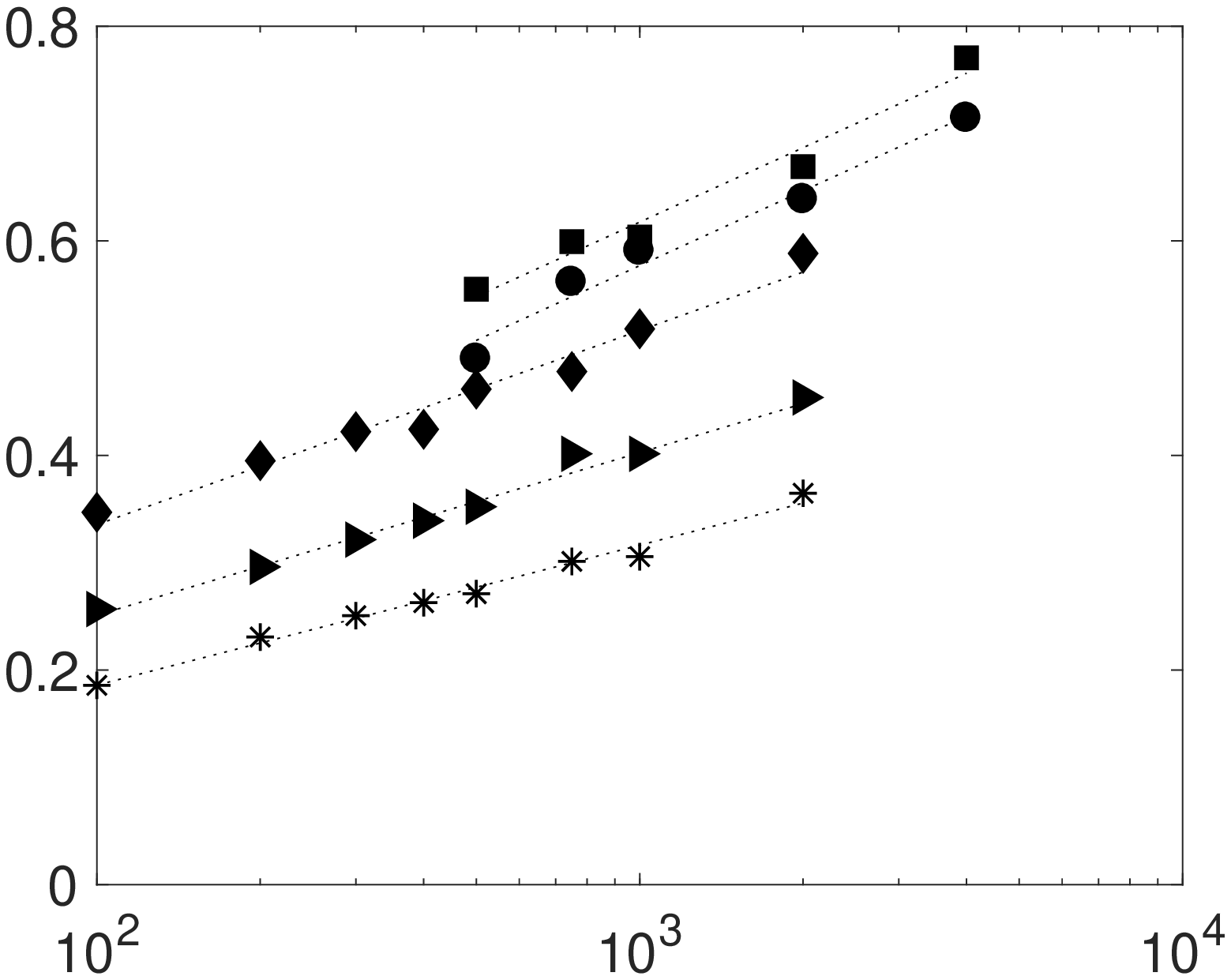}}
        \put(180,25){\includegraphics[width = 0.46\textwidth]{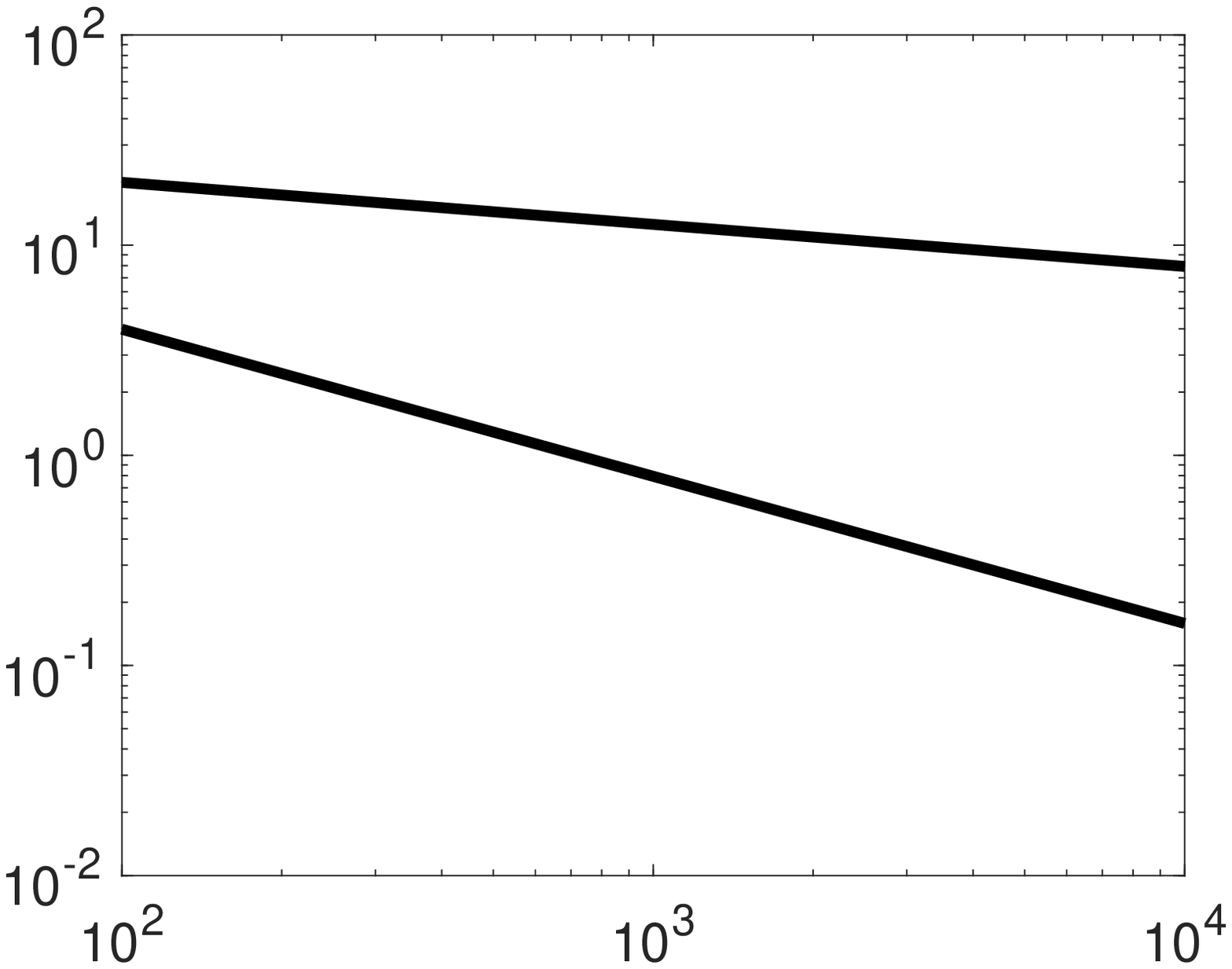}}
        \put(-10,95){$\alpha_2(N)$}
        \put(5,235){$m$}
        \put(175,235){$m$}
        \put(180,95){$m$}
        \put(95,17){$N$}
        \put(95,162){$N$}
        \put(260,162){$N$}
        \put(260,17){$N$}
       % \put(89,145){$(a)$}
       % \put(249,145){$(b)$}
       % \put(89,0){$(c)$}
        %\put(249,0){$(d)$}
        \put(220,60){$\textrm{\it connected}$}
        \put(250,103){$\textrm{\it clustered}$}
        \put(265,130){$\textrm{\it disconnected}$}
    \end{picture}\vspace*{-5mm}
    \caption{Top: scatter plots of triangle density $m$ shown versus system size $N$. The width of the markers is proportional to the number of connected components. Top left: $p(k) = bim(k|3,7)$, with dashed-dotted line corresponding to the prediction $p(q)q(q-1)$ of (\ref{eq:malphaK}). The observed slope of $-0.64(5)$ is consistent with the predicted $-0.625$ of (\ref{eq:mu-upper-bound}). Top right: $p(k) = PL(k)$, and inset $p(k) = exp(k|5)$. Here solid lines are only guides to the eye. Bottom left: linear/log plot of $\alpha_2(N)$ versus $N$ from simulation data. Dotted line shows linear fit in good agreement with theoretical prediction (\ref{eq:alpha2}), see table \ref{tab:comparison}. Bottom right: conjectured phase diagram of the ensemble (\ref{eq:ensemble}), in the $(m,N)$ plane.}
    \label{fig:phaseDiagram}
\end{figure}

With our estimate for $\alpha_2(N)$ we can also derive an upper bound on the loop density achieved in the connected regime,
\begin{align}
    \label{eq:mu-upper-bound}
    m_u = m_t(\alpha_2(N)) =\frac{1}{N^{\frac{q-2}{q+1}}} (\overline{k^2}/c - 1)^3 (p(q)q(q-1)/c_q)^{\frac{6}{q(q^2-1)}}
\end{align}
This value corresponds to the loop density that would be reached if the contribution of cliques were not present, given that cliques appear before it becomes impossible to reach this density in the connected phase. Even though $c_q$ is unknown,  we can conclude that $m_u$ vanishes when $N\to\infty$, which  is indeed consistent with numerical experiments, as can be seen in Figure \ref{fig:phaseDiagram}. The results are very good when looking at the chosen bimodal degree distributions, $p(k) =\frac{1}{2}\delta_{k3}+\frac{1}{2}\delta_{kq}$. Figure \ref{fig:phaseDiagram} confirms two theoretical predictions.  First, we see that the last value of the loop density before the steep jump into the clustered phase scales with $N$ in the manner  predicted by (\ref{eq:mu-upper-bound}). Second, the final value of the jump at $\alpha_2(N)$ coincides with the prediction $p(q)q(q-1)$, as indicated by the dotted-dashed line in Figure \ref{fig:phaseDiagram}.

As a final comment, we point out that the Poissonian assumption of (\ref{eq:poissonian}) implies that the shattering transition is of an entropic nature. To see this, we can study the behaviour of the ratio 
\begin{align}
    \frac{\calN(\bk|T)}{\calN(\bk|K)} = \frac{\# \textrm{of graphs with degree sequence }\bk\textrm{ and }T\textrm{ isolated triangles} }{\# \textrm{of graphs with degree sequence }\bk\textrm{ and }K\textrm{ q-regular cliques}}
\end{align}
If we fix the loop density to any arbitrary value $m^*<p(q)q(q-1)$, this value can be achieved by the following  numbers of triangles or cliques.
\begin{align}
    T =  \frac{m^*}{6}N,~~~~~~
    K =  \frac{m^*}{q(q-1)}N
\end{align}
Using the Poissonian assumption,  we can then prove (see  \ref{sec:entropic}) that
\begin{align}
    \lim_{N\to\infty} \frac{\calN(\bk|m^*N/6)}{\calN(\bk|m^*N/(q^2-q))} = \lim_{N\to\infty}\rme^{-\frac{m^*}{6}\frac{q-2}{q+1}N\log N} = 0
\end{align}
Hence, no matter how  small $m^*$ is, for a large enough system there will always be infinitely many more graphs that achieve it via cliques than via isolated triangles.

\begin{table}[t]
    \centering
    \caption{Comparison of the slope of $\alpha_2(N)$ plotted against $\log N$, as measuerd from data in  Figure \ref{fig:phaseDiagram}, versus the theoretically predicted value $[2(q+1)]^{-1}$ of (\ref{eq:alpha2}). The degree distributions  $bim(k,a,b)$, $u(k)$ and $v(k)$ are defined as in Table 1.}
    \begin{indented}
    \item[]\begin{tabular}{@{}llllll}
    \br
    $p(k)$&$bim(k|3,5)$&$bim(k|3,7)$&$bim(k|3,9)$&$u(k)$&$v(k)$\\
    \mr &&&&&\\[-1em]
    theory     &$0.08\Bar{3}$&$0.0625$&$0.05$&$0.08\Bar{3}$&$0.08\Bar{3}$  \\
    &&&&&\\[-1em]
    simulation &0.079(5)&0.066(2)&0.057(3)&0.10(1)&0.10(1)\\
    \br
    \end{tabular}
    \end{indented}
    
    \label{tab:comparison}
\end{table}

\section{Discussion}

In this letter we have presented and analyzed a random graph ensemble were samples are both sparse and loopy. Even though this ensemble (\ref{eq:ensemble}) can be regarded as the simplest random loopy graph ensemble,  it is found to exhibit rather nontrivial behaviour. While one would hope for and expect a smooth and easy controllability of the loop density via the control parameter $\alpha$, we see that in fact there are very special nontrivial regimes, and there is surprisingly a very strong influence of the system size, i.e. the number of nodes in the graphs. 
Still, with appropriate care this ensemble could be used by practitioners of network science as a null model of loopy networks. If one has a given real network $\bA_0$, that is to be compared with random samples having the same loop density $m(\bA_0)$,  we propose the following steps should be taken:
\begin{enumerate}
    \item Calculate the following properties of the initial graph: $\bk(\bA_0), m(\bA_0), r(\bA_0), n(\bA_0)$
    \item Sample graphs repeatedly from (\ref{eq:ensemble}),  varying $\alpha$ until the value $\alpha^*$ where observed and required loop densities match, $m(\alpha) = m(\bA_0)$. An initial guess for $\alpha$ might be $\alpha_0 = \frac{1}{6}\log(m(\bA_0)N/(\overline{k^2}/c-1)^3)$, especially if if $\alpha_1(N)>\alpha_0$.
    \item Once loop densities are matched, compare the other properties $r(\bA)$ and $n(\bA)$. 
    \begin{itemize}
        \item If $n(\alpha)\approx n(\bA_0)$ and $r(\alpha)  \approx r(\bA_0)$, then (\ref{eq:ensemble}) is a suitable null model for $\bA_0$.
        \item If they are different, it means that $\bA_0$ is still extremely atypical in (\ref{eq:ensemble}), and thus it is not a suitable null model
    \end{itemize}
\end{enumerate}
Even if all observables $m(\bA)$, $r(\bA)$ and $n(\bA)$ of initial and sampled graphs match, it still could be the case that equilibration waiting times of the MCMC are very large. For graphs of more than a thousand nodes it could take days or more to get well-mixed samples. This just shows how the applied network scientist should be cautious when applying tools like edge swapping without a proper theory. 

To summarize, we present our conjectured phase diagram in Figure \ref{fig:phaseDiagram} (bottom right). With an exact solution for (\ref{eq:freeEnergy}) one could find an analytic expression for the phase boundaries shown. The main lessons are that the same loop densities may have very different topologies for different systems sizes, and that sampling anywhere outside the connected regime takes a very long time, potentially days or weeks for large graphs, even on fast multi-core machines. We expect that for any model, any desired loop density eventually falls in the disconnected regime as $N$ grows. For the case of bounded degree distributions with $Np(q)>>q+1$, the clustered region practically vanishes.

There are many directions in which to pursue further research, ranging from practical to theoretical. From a rigorous point of view it would be interesting to see how to prove or disprove any of the assertions made in this work, that is extending rigorous results of CM beyond uniform models. Additionally, longer and more extensive simulations should be carried out to try to determine the exact dependence on $N$ of $\alpha_1(N)$ and $\alpha_2(N)$, especially to find out whether there is indeed a transition without scaling parameters for unbounded degree distributions.

The enormous waiting times seem to be due in part to the fact that in the clustered and disconnected phases many loops have to broken in a predetermined sequence to get rid of certain structures like cliques. Given that this is unlikely, an alternative MCMC with moves that involve more edges rather than only $2$ could be studied, in order to speed up the algorithm and let it explore more quickly the graph space.

Finally, there are many interesting questions about the spectral properties of (\ref{eq:ensemble}) to discover. First, in \cite{lopez2020imaginary} an analytic expression for the spectral density was found for the case of regular graphs in the connected regime. We are currently working on a generalization for an arbitrary degree distribution like in (\ref{eq:ensemble}). The formation of clusters after the clustering transition points to a localization transition for the eigenvectors of $\bA$. A similar observation has been made for dense graphs in \cite{avetisov2019localization}, where its nontrivial spectral properties were found; such spectral analysis has not been done yet for the sparse case like ours.

Overall, there are many open question when it comes to presenting random counterparts of real networks. It is safe to say that they are not defined by loopiness alone. It seems like real networks occupy a very small area of the abstract graph space. Finding the correct properties that will make a maximum entropy ensemble sample from a pool of realistically looking graphs is still very much an open problem. An alternative is to impose a constraint on the full set of eigenvalues of the adjacency matrix, in this way all loop lengths would be controlled simultaneously. This full spectral constraint has been discussed in \cite{CoolenLoopy,roberts2014random,lopez2020imaginary}.
\\[3mm]
{\bf Acknowledgements}
\\[1mm]
FAL gratefully acknowledges financial support through a scholarship from Conacyt (Mexico). The authors thank Alexander Mozeika and Mansoor Sheikh for valuable discussions.

\section*{References}
% \bibliography{bibl}

\providecommand{\newblock}{}

\appendix

\section{Numerical sampling\label{sec:numerical-sampling}}

In order for this paper to be sufficiently self-containment, we will present a brief recap of the algorithms described in \cite{annibale2017generating,sampling} for generating  samples from nondirected random graph ensembles with hard-constrained degrees. The main task is to define a Markov chain with the following characteristics:
\begin{align}
    p_{t+1}(\bA) = \sum_{\bA' \in \Omega_{\calM}} W(\bA |\bA^\prime)p_t(\bA^\prime),
\end{align}
\begin{enumerate}
    \item The measure $p_t$ converges to the invariant measure $p_{\infty}(\bA) = \frac{1}{Z} \rme^{-H(\bA)}$. 
    \item The allowed transitions constitute a limited set $\Phi$ of elementary moves \begin{align*}
    F:~\Omega_F \subseteq \Omega_{\calM} \to \Omega_{\calM}
    \end{align*}
    \item For each $F\in\Phi$ there exists a unique inverse $F^{-1}$ that acts on the same set of graphs, $\Omega_{F^{-1}} = \Omega_F$
\end{enumerate}
\begin{figure}[t]
    \centering
    \includegraphics[width=.7\textwidth]{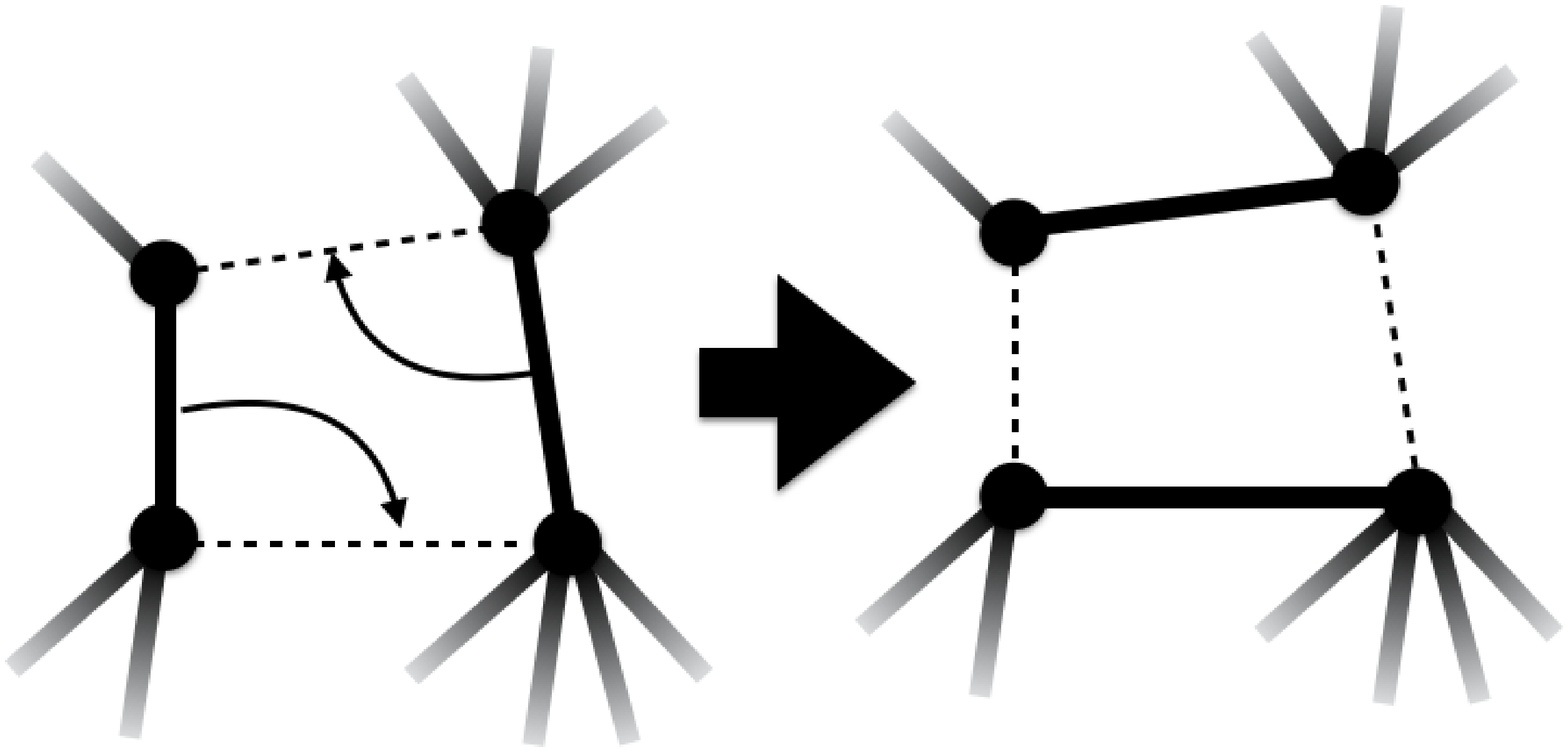}
    \caption{Edge swap for MCMC dynamics in the space of simple nondirected graphs. This is the simplest type of move in  that leaves all degree invariant. }
    \label{fig:edgeswap}
\end{figure}
With these condition we will be able to define a dynamical process that will allow us to sample effectively from ensemble (\ref{eq:ensemble}). The reason we need nontrivial moves is to be sure we respect the degree constraints; a single edge dynamics cannot achieve this. The simplest elementary move that respects the values of all degrees  is called an edge swap. It involves choosing a pair of edges and interchanging them, see Figure \ref{fig:edgeswap}.

We next need to define the transition probabilities $W(\bA |\bA^\prime)$ of the Markov chain. They are chosen such as to obey the detailed balance condition, with (\ref{eq:ensemble}) as invariant measure, i.e. $W(\bA|\bA^\prime) p_{\infty}(\bA^\prime) = W(\bA^\prime|\bA) p_{\infty}(\bA)$ for all $(\bA,\bA^\prime)$. Together with the known ergodicity of the edge swap moves \cite{eggleton1981simple}, detailed balance is a sufficient condition to satisfy (i). We can write the transition probabilities as
\begin{align}
    W(\bA|\bA') = \sum_{F\in\Omega'}\frac{I_{F}(\bA')}{n(\bA')}\s{\delta_{\bA,F\bA'} A(F\bA'|\bA') + \delta_{\bA,\bA'}\s{1-A(F\bA'|\bA')}}.
\end{align}
with the definitions
\begin{align}
    \Omega' &= \left\{F\in\Phi |~\exists \bA\in\Omega_{\calM} \textrm{ s.t. }F\bA\not = \bA\right\}\nonumber\\
    I_F(\bA) &=\left\{\begin{array}{cc}
    1     &\textrm{ if }\bA\to F\bA \textrm{   is an allowed move}  \\
    0     & \textrm{ otherwise}
    \end{array}\right.\nonumber\\
    n(\bA) &=\sum_{F\in\Omega'}I_F(\bA)\nonumber\\
    A(F\bA|\bA) &:\textrm{ acceptance probability of move }\bA\to F\bA
\end{align}
The interpretation of the above transition probabilities is as follows. At each step a candidate move is chosen uniformly at random from all possible moves, with probability $1/n(\bA)$. It is then accepted with probability $A(F\bA|\bA)$, and otherwise rejected. The acceptance probabilities must satisfy the detailed balance condition
\begin{align}
    (\forall\bA\!\in\!\Omega)(\forall F\!\in\! \Omega'):\:\: A(F\bA|\bA)\rme^{-H(\bA)}/n(\bA) = A(\bA|F\bA)\rme^{-H(F\bA)}/n(F\bA)
\end{align}
This condition is satisfied by multiple choices; here we choose
\begin{align}
    A(\bA|\bA') =  \frac{1}{1+\rme^{E(\bA)-E(\bA')}}
\end{align}
with the effective energy $E(\bA) = H(\bA) + \log n(\bA)$.  This expression stresses the fact that the acceptance probabilities cannot depend only on the function $H(\bA)$, but also on the current state via $n(\bA)$. In \cite{annibale2017generating} it is shown that $n(\bA)$ an be written explicitly as
\begin{align}
    n(\bA) = \frac{1}{4}\Big(\sum_i k_i\Big)^2 + \frac{1}{4}\sum_i k_i - \frac{1}{2}\sum_i k_i^2 - \frac{1}{2}\sum_{ij}k_i A_{ij} k_j +\frac{1}{4}\Tr\p{\bA^4} + \frac{1}{2}\Tr\p{\bA^3}
\end{align}

\section{Entropic argument \label{sec:entropic}}

Let us assume that in the configuration model (CM) both $T$ and $K$ are Poissonian random variables,
\begin{align}
    P_N(T) =  Poiss(T|\lambda_t),~~~~~~
    Q_N(K) = Poiss(K|c_q/N^{d-1})
\end{align}
with $d=\frac{1}{2}q(q-1)$. 
They are simply related to the number of graphs that exist, given the prescribed degree sequence, with the stated number of triangles or cliques, so
\begin{align}
    \frac{P_N(T)}{Q_N(K)} = \frac{\sum_\bA \delta_{T,T(\bA)}\prod_{i=1}^N\delta_{k_i,\sum_j A_{ij}}}{\sum_\bA \delta_{K,K(\bA)}\prod_{i=1}^N\delta_{k_i,\sum_j A_{ij}}} = \rme^{-\lambda_t + \frac{c_q}{N^d}}\frac{(\lambda_t)^T}{(c_q/N^{d-1})^K}\frac{K!}{T!}
\end{align}
We want to determine for a given loop density whether asymptotically  there are  more graphs that realize the joint values $(T,K)$   through triangles or through cliques. For this we need to write the number of triangles and cliques in terms of the desired loop density, which gives
\begin{align}
    T =  \frac{m}{6}N,~~~~~~
    K =  \frac{m}{q(q^2-1)}N
\end{align}
We can now  inspect the asymptotic limit
\begin{align}
    \lim_{N \to \infty} \frac{P_N\p{\frac{m}{6}N}}{Q_N\p{\frac{m}{q(q^2-1)}N}} = \lim_{N\to\infty} & \exp\left(-\lambda_t + \frac{c_q}{N^d} + \frac{m}{6} N \log \lambda_t - \frac{m}{q(q^2-1)}N\log(c_q) \right.\nonumber\\
    & \left. + \frac{m d}{q(q^2-1)} N \log(N) + \p{\frac{m}{q(q^2-1)}N}! - \p{\frac{m}{6}N}! \right)
\end{align}
We note, upon using Stirling's expression for the factorials, that this quantity is dominated by the $N\log N$ term, since $d = \frac{1}{2} q(q-1)$. Hence
\begin{align}
    \lim_{N\to\infty}\frac{P_N\p{\frac{m}{6}N}}{Q_N\p{\frac{m}{q(q^2-1)}N }} = \lim_{N\to\infty} \exp\p{  - m\p{\frac{1}{6} -\frac{1}{2(q+1)} } N \log N} = 0.
\end{align}
\end{document}